\documentclass[twocolumn,showpacs,preprintnumbers,amsmath,amssymb]{revtex4} 

\usepackage{graphicx,color}
\usepackage{dcolumn}
\usepackage{bm}

\begin{document}
\title{Comparing proteins by their internal dynamics: exploring structure-function relationships beyond static structural alignments}

\author{Cristian Micheletti\\
$\null$\\
{Scuola Internazionale Superiore di Studi Avanzati, via Bonomea 265, Trieste, Italy;\\
e-mail: {\tt michelet@sissa.it}}}
\date{\today}

\begin{abstract}
The growing interest for comparing protein internal dynamics owes much
to the realization that protein function can be accompanied or assisted
by structural fluctuations and conformational changes.  Analogously to
the case of functional structural elements, those aspects of protein
flexibility and dynamics that are functionally oriented should be
subject to evolutionary conservation. Accordingly, dynamics-based
protein comparisons or alignments could be used to detect protein
relationships that are more elusive to sequence and structural
alignments. Here we provide an account of the progress that has been
made in recent years towards developing and applying general methods for
comparing proteins in terms of their internal dynamics and advance the
understanding of the structure-function relationship.\\
\phantom{XX}\\
{\small Link to published article in Physics of Live Reviews: http://dx.doi.org/10.1016/j.plrev.2012.10.009}
 \end{abstract}

\maketitle

\section{Introduction}

Over the past decades enormous efforts have been made to clarify the
sequence $\to$ structure $\to$ function relationships for proteins and
enzymes. In particular the sequence $\to$ structure connection has been
extensively probed by dissecting the detailed physico-chemical
mechanisms that assist and guide the folding process of several
proteins~\cite{creighton,finkelbook,lesk,fersht}. The more general
aspects of this relationship are, however, better captured by analysing
the degenerate mapping between the ensembles of naturally-occurring
protein sequences and their corresponding
folds\cite{Chothia:1986:EMBO-J:3709526,chothia,lesk,Chothia:1990:Annu-Rev-Biochem:2197975,holm96,Holm:2000:Bioinformatics:10980157,fleming,Meyerguz:2007:Proc-Natl-Acad-Sci-U-S-A:17596339}.
For instance, the current $\sim$85,000 entries can be clustered in about
20,000 non-redundant sequence sets
but cover only 1,500 distinct structural folds\cite{SCOP,CATH}.

The introduction of general quantitative schemes for comparing, or
aligning, protein sequences and protein structures has played a crucial
role for framing the observed many-to-one sequence-structure
relationship in the context of molecular
evolution\cite{Orengo:2005:Annu-Rev-Biochem:15954844,Ranea:2006:J-Mol-Evol:17021929}. In
particular, by following the impact that evolutionary sequence
divergence has on native structural changes
\cite{Chothia:1986:EMBO-J:3709526} it has been possible to identify
general properties of peptide chains, amino acid hydrogen-bonding
patterns, thermodynamic stability etc. that govern the
sequence-structure relationship by constraining the repertoire of viable
structural changes that are evolutionary
accessible\cite{Chothia:1990:Annu-Rev-Biochem:2197975,Williams:2009:Mol-Biol-Evol:19193735,PhysRevLett.84.3009,PhysRevLett.82.3372,Maritan:2000:Nature:10917526,DePristo:2005:Nat-Rev-Genet:16074985,ranganathan03,Shakhnovich:2005:Genome-Res:15741509,Zeldovich:2008:Annu-Rev-Phys-Chem:17937598}.

As a result, remote evolutionary relationships are more confidently
obtained from structure-based comparative methods than sequence based
ones.

Besides the above general constraints, additional and stronger ones are
imposed by functional requirements. In fact, it has long been known that
enzymes that have evolutionarily diverged and that catalyze different
reactions, tend to conserve very precisely functional structural
elements and the location of the active site where different amino acids
can be recruited for different
function\cite{Chothia:1986:EMBO-J:3709526,bartlett,Weinreich:2006:Science:16601193,Ojha:2007:PLoS-Comput-Biol:17658942,Pegg:2006:Biochemistry:16489747}. More
recently it has also emerged that specific features of protein internal
dynamics that impact biological activity and functionality can also be
subject to evolutionary
conservation\cite{cascella,Zheng:2005:Structure:15837195,NM_GROEL,Ramanathan:2011:PLoS-Biol:22087074,Liu:2012:Mol-Biol-Evol:22427707}.

By analogy with the sequence-structure case, one may therefore envisage
that quantitative methods apt for comparing function-oriented properties
in different proteins could advance the capability of detecting protein
evolutionary relationships that may be elusive to sequence- or
structure-based investigations.

Here we shall review recent studies which focused on the comparison of
protein internal dynamics, which is arguably one of the many aspects
that often, though not always, assist or influence protein function over
a wide range of time
scales\cite{eisenmesser,Min:2005:Phys-Rev-Lett:16090221,Boehr:2008:Biochemistry:18690714}. For
example, concerted structural movements in enzymes, either ``innate'' or
triggered by ligand binding, have been argued to be important for
enzymes to achieve a catalytically-competent state, promote catalytic
efficiency, for allosteric signal propagation and protein-protein
interactions\cite{eisenmesser,HenzlerWildman:2007:Nature:18026087,HenzlerWildman:2007:Nature:18075575,Ramanathan:2011:PLoS-Biol:22087074,piana02b,micheletti04,Nashine:2010:Curr-Opin-Chem-Biol:20729130,Agarwal:2002:Proc-Natl-Acad-Sci-U-S-A:11867722,Morra:2009:PLoS-Comput-Biol:19300478,Stein:2011:Structure:21645858,delSol:2009:Structure:19679084,Liu:2012:Mol-Biol-Evol:22427707,HammesSchiffer:2006:Annu-Rev-Biochem:16756501,Zheng:2009:Curr-Protein-Pept-Sci:19355980,Miyashita:2005:J-Phys-Chem-B:16851180,Miyashita:2003:Proc-Natl-Acad-Sci-U-S-A:14566052,Chennubhotla:2007:PLoS-Comput-Biol:17892319,Vishveshwara:2009:Curr-Protein-Pept-Sci:19355982,McCammonAIDS,NM_GROEL,Provasi:2011:PLoS-Comput-Biol:22022248,Teilum:2009:Cell-Mol-Life-Sci:19308324,Bavro:2012:Nat-Struct-Mol-Biol:22231399,Eisenmesser:2005:Nature:16267559,Bhabha:2011:Science:21474759,Jackson:2009:Proc-Natl-Acad-Sci-U-S-A:19966226,Glembo:2012:PLoS-Comput-Biol:22479170,Zen:2010:BMC-Struct-Biol:20691107}.

We shall accordingly report on the progress that has been made in recent
years towards developing and exploiting quantitative numerical
strategies for comparing the internal dynamics of proteins and explore
its connection with structural and functional similarities.

The material presented in the review is organised as follows.  Because
these approaches are virtually all based on numerical characterizations
of protein internal dynamics we shall first provide a self-contained
methodological summary of the theoretical/computational techniques used
to characterize and compare protein internal dynamics.  Next we shall
overview the contexts where dynamics-based comparisons, with different
resolution and scope, have been applied. We shall further provide an in
depth discussion of a number of selected instances where dynamics-based
similarities have been detected within structurally-heterogeneous
members of specific protein families, and even across protein families.

\section{Comparing protein internal dynamics: methodological aspects}

In this section we provide a self-contained overview of the quantitative
numerical approaches employed to characterize and compare the internal
dynamics of proteins.  In particular, we first review the essential
dyamics analysis techniques which are commonly applied to atomistic
molecular dynamics simulations or phenomenological coarse-grained models
(elastic networks) to single out the collective degrees of freedom that
best account for protein's internal motion in thermal equilibrium.  Next
we shall discuss how the essential dynamical spaces and other
dynamics-related quantities can be used for comparative purposes.

\subsection{Protein internal dynamics: essential dynamics analysis of MD trajectories}
\label{sec:ed}

The wealth of information produced by extensive atomistic molecular
dynamics (MD) simulations of globular proteins is typically described
and rationalised by identifying the few collective degrees of freedom
that best capture the internal protein dynamics. Arguably, the most
commonly used technique is represented by the principal component
analysis\cite{garcia92} of amino acid pairwise displacements.

This technique relies on the spectral decomposition of the matrix of
pairwise correlations of  the displacements of amino acids, represented by their $C_\alpha$ atoms,  from their reference positions.

In the following we shall indicate with ${\bf r}_i(t)$ the three-dimensional position at simulation time $t$ of the $i$th  $C_\alpha$
atom and with $\delta{\bf r}(t) \equiv {\bf r}_i(t) - \langle {\bf r}_i\rangle$ the associated vector displacement from the
average reference position. A generic entry of the matrix of pairwise displacement correlations, $C$, is accordingly defined as

\begin{equation}
C_{ij,\mu\nu} = \langle \delta r_{i,\mu}(t) \delta r_{j,\nu}(t) \rangle \,
\label{eqn:c}
\end{equation}

\noindent where $\delta r_{i,\mu}(t)$ is the $\mu$th Cartesian component of the vector displacement of the $i$th amino acid and  $\langle\rangle$ denotes the average over simulation time. For proteins consisting of $N$ amino acids, the symmetric covariance matrix $C$ has linear size equal to $3N$.

It is important to notice that the matrix element of eq.~\ref{eqn:c} can be equivalently rewritten as:
\begin{equation}
C_{ij,\mu\nu} = \sum_{l=1}^{3N} \lambda_l \, v^l_{i,\mu} v^l_{j,\nu}
\label{eqn:spectr}
\end{equation}

\noindent where $\lambda_1,\, \lambda_2,...$ are the eigenvalues of $C$
ranked by decreasing magnitude and ${\bf v}^1,\, {\bf v}^2,...$ are the
corresponding orthonormal eigenvectors.  

Because the protein overall mean square fluctuation is given by
\begin{equation}
\sum_{i,\mu} \langle \delta r_{i,\mu}(t)^2 \rangle = \sum_{i,\mu} C_{ii,\mu\mu}= \sum_l \lambda_l
\end{equation}
\noindent one has that top ranking eigenvectors of $C$ embody the independent degrees of freedom that most contribute to the internal dynamics of the protein.
Indeed, for most globular proteins of 100-200 amino acids, the top 10
eigenvectors suffice to capture most of the protein mean square fluctuation\cite{garcia92}. For this reason, considerations are typically restricted to the linear space spanned by the top eigenvectors of $C$, which is commonly termed the essential dynamical space\cite{amadei93}.

The structural deformations entailed by the essential eigenvectors, or
essential modes, are typically found to embody concerted, collective
displacements of protein subportions consisting of several amino
acids\cite{garcia92,tirion}.  As a matter of fact, the large-scale
collective conformational changes that many proteins and enzymes need to
sustain in order to carry out their biological functionality have been
shown to lie in the essential dynamical
space\cite{Tama:2001:Protein-Eng:11287673,sanejouand,falke,rod,gerstein,wall,smith,NM_GROEL,Pontiggia:2008:Biophys-J:18931260}.

These observations provide an {\em a posteriori} justification for
considering the essential dynamical spaces as providing key information
into functionally-oriented aspects of proteins.

We conclude by noting that one relevant technical point of the essential
dynamics analysis regards the definition of the reference amino acids
positions from which the instantaneous displacements $\delta {\bf r}$
are calculated. For proteins that have an overall rigid-like character,
these positions can be obtained by averaging the conformers sampled by
the MD simulation after optimally superposing them. The structural
superposition is necessary to remove the overall rotations and
translations of the molecules. It is important to stress that this step
is not trivially accomplished when proteins have an appreciable internal
flexibility character (e.g. due to the presence of mobile subdomains)
\cite{Zhou:2000:Biophys-J:11106598}. In this case, to avoid artefactual
results, it is crucial to identify the correct frame of reference for
describing and computing the internal structural fluctuations of the
protein, see e.g. the discussion of
ref.~\cite{HenzlerWildman:2007:Nature:18026086,HenzlerWildman:2007:Nature:18026087}
and related supporting material.

However, it must be noted that the relative displacements of domains in
multidomain proteins can be so large that protein movements cannot be
reliably described by a linear superposition of a limited number of
essential dynamical spaces, even if obtained with the above-mentioned
procedure.  A prototypic example is offered by the relative rotation of
protein domains by a finite angle. In this case the directions of
instantaneous rotations of the two extreme positions can project very
poorly on the difference vector of the latter (see Fig. 3 in
ref. \cite{Song:2006:Proteins:16447281}). In such cases the salient
degrees of freedom of protein internal dynamics can be identified by
decomposing the protein of interest into quasi-rigid
domains\cite{hinsen_domains,yesy,Camps:2009:Bioinformatics:19429600,dyndom,kundu,thorpe,Potestio:2009:Biophys-J:19527659,Aleksiev:2009:Bioinformatics:19696046}
and next considering their relative roto-translations
\cite{Morra:2012:PLoS-Comput-Biol:22457611}, see also section \ref{sec:multidom}.

\subsection{Essential dynamical spaces from elastic network models}
 
The collective character of the top eigenvectors of the covariance
matrix $C$ obtained from atomistic MD simulations suggests that the
essential dynamical spaces could be reliably identified by
coarse-grained protein models.

This observation, which was stimulated by the seminal work of M. Tirion
\cite{tirion} has in fact lead
to the introduction of the well-known elastic network models which,
despite adopting a simplified description of a protein's structure and
its native amino acid interactions, can reliably identify the essential
dynamical spaces of globular proteins with a negligble computational
expenditure\cite{bahar97,Hinsen98,atilgan,sanejouand,Micheletti:2001:Phys-Rev-Lett:11497984,micheletti04,gfpgauss}.

In these approaches, each amino acid is described by one or few
centroids (e.g. the C$_\alpha$ atom for the main chain \cite{atilgan}
and an additional centroid for the side chain\cite{micheletti04}) the
model potential energy is constructed by introducing quadratic penalties
for the deviations from the native values of the distance of all pairs
of centroids that are in contact in the native state. 
Accordingly, for a protein consisting of $N$ amino acids, the resulting potential energy
has the form:
\begin{equation}
U= {1 \over 2} \sum_{ij,\mu\nu} \delta r_{i,\mu} \, M_{ij,\mu\nu} \delta r_{j,\nu} .
\label{eqn:U}
\end{equation}

\noindent where $M$ is a symmetric matrix of linear size $3N$.  In the
following we shall indicate with $\tau_0$, $\tau_1$,... $\tau_{3N}$ the
eigenvalues of $M$ ranked for {\em increasing} magnitude, and with ${\bf
  w}^0$, ${\bf w}^1$,... ${\bf w}^{3N}$ the associated orthonormal
eigenvectors.  The eigenvalues $\{\tau_l\}$ are all positive except for
the six attributed to the global rotations and translations of the
molecule. It is evident that eq.~\ref{eqn:U} bears strong analogies with
the normal mode analysis of proteins\cite{levitt85,tirion}.

Because of the quadratic character of the model potential energy of eq. (\ref{eqn:U})
canonical equilibrium properties of the elastic network can be
calculated exactly. In particular, a generic entry of the model covariance
matrix $C$ is given by
\begin{equation}
C_{ij,\mu\nu} =  \kappa_B T \,  \tilde{M}^{-1}_{ij,\mu\nu} 
\end{equation}

\noindent where $\kappa_B T$ is the thermal energy at the temperature of
interest, $T$, and the tilde superscript denotes the pseudoinversion
operation i.e. the removal of the zero-eigenvalue space prior to the
inversion of $M$. Equivalently, $C$ can be written as
\begin{equation}
C_{ij,\mu\nu} = \sum_l^\prime {\kappa_B T \over \tau_l} \, w^l_{i,\mu} w^l_{j,\nu}
\label{eqn:Cenm}
\end{equation}
\noindent where the prime indicates the omission of the eigenspaces
associated to the zero eigenvalues. The above expression clarifies that
the degrees of freedom that most account for the proteins' fluctuations
in thermal equilibrium correspond to the modes of protein deformation
associated to the smallest eigenvalues, i.e. those that cost least
energy to excite. 

If the proteins dynamics were described by an overdamped Langevin
scheme, these low-energy modes would also be those having the slowest
relaxation time. Although the harmonic character of the near-native free
energy well and the white noise Langevin description apply only
limitedly to proteins
\cite{Karplus76,BrooksI,BrooksII,Min:2005:Phys-Rev-Lett:16090221,Pontiggia2007,Hinsen:2008:Proteins:17853448},
the observation is qualitatively consistent with the fact that
collective low-energy modes in proteins occur over long time scales (and
hence are occasionally referred to as ``low-frequency'' modes).
These observations motivate the practice, adopted in this review too, of
regarding the principal components of equilbrium structural fluctuations
as embodying the salient internal dynamical properties.

We conclude by mentioning that in recent years alternative formulations
of elastic network models have been proposed including versions based on the
matching of observables obtained from atomistic MD
simulations\cite{orozco2010} and on the use of internal coordinates, which are commonly used in normal mode analysis too\cite{Go:1983:Proc-Natl-Acad-Sci-U-S-A:6574507,doi:10.1021/ct050307u,Mendez:2010:Phys-Rev-Lett:20867208,Orozco:2011:Adv-Protein-Chem-Struct-Biol:21920324,LopezBlanco:2011:Bioinformatics:21873636}.

\subsection{Anharmonicity of proteins free energy landscape}

The viability of elastic network models to capture the salient traits of
protein conformational fluctuations is justified {\em a posteriori} by
the good accord between the essential covariance matrices of elastic
network models and of extensive atomistic MD simulations.  For example
in ref.~\cite{micheletti04} it was compared the covariance matrices of
HIV-1 protease with a bound ligand obtained from a 14-ns MD simulation
with an atomistic force-field and explicit solvent and the beta-Gaussian
elastic network model, which employs two centroids per amino acids (for
main- and side-chain, respectively). The linear correlation coefficient
of the $\sim$20,000 corresponding distinct entries of the two matrices
was significant (equal to 0.8) like the consistency of the two sets
of essential dynamical spaces.
{A more recent example of the good accord of protein structural fluctuations computed with elastic network models and MD atomistic simulations is provided by the work of Romo and Grossfield on GPCRs membrane proteins\cite{Romo:2011:Proteins:20872850}. This study showed that a suitably-parametrized model can match the essential dynamical spaces and their relative weight observed in microsecond-long simulations\cite{Romo:2011:Proteins:20872850}.}

This agreement is noteworthy in consideration of the highly
complex free energy landscape explored by folded proteins can explore in
thermal equilibrium. In fact, this landscape presents several tiers of
local minima
\cite{Frauenfelder:1991:Science:1749933,Wolynes:1995:Science:7886447,Frauenfelder:1988:Annu-Rev-Biophys-Biophys-Chem:3293595}
with low barriers (compared to the thermal energy $\kappa_B T$)
separating conformational states with local structural differences such
as the rotameric state of a sidechain while large ones separate
conformational ensembles with major subdomains rearrangement, such as
for open and closed conformation of certain enzymes. In turn, the
hierarchical organization of these minima reflects in a broad range of
time-scales, from the ps to the ms and beyond, over which the mentioned
structural changes can occur as observed in NMR and single-molecule
experiments \cite{HenzlerWildman:2007:Nature:18026087,HenzlerWildman:2007:Nature:18075575,Boehr:2008:Biochemistry:18690714,yang,Min:2005:Phys-Rev-Lett:16090221}

From these general considerations and from the detailed analysis of the
protein conformational substates visited over MD trajectories of
hundreds of ns\cite{Pontiggia:2008:Biophys-J:18931260,Ramanathan:2011:PLoS-One:21297978}
it emerges that the harmonic approximation on which elastic network
models rely may be a highly simplified parametrization of even the
near-native free energy landscape.

While this limitation, that may be more or less severe depending on the
molecule rigidity, must be clearly be borne in mind, it should be noted
that the free-energy landscape of a few proteins has been shown to be
endowed with particular properties that make the harmonic, or
quasi-harmonic
\cite{McCammon.BIOP.1984,BrooksIII,Hayward:1994:Protein-Sci:7520795,hinsen}
free-energy approximation still informative even when dealing with major
and slow conformational changes.  Specifically, computational studies of
lysozyme \cite{Kitao:1998:Proteins:9849935}, protein
G\cite{Pontiggia2007} and adenylate kinase
\cite{Pontiggia:2008:Biophys-J:18931260} has {clarified} that the
principal directions of the free energy minima associated to the
substates populated by each of these proteins are very consistent with
each other and also very similar to the difference vectors connecting
the substates themselves. This indicates that, despite their structural
differences, different substates of the same protein tend to have very
similar modes of conformational fluctuations and that the latter, in
turn, predispose the observed conformational changes between substates.
Indeed, by analyzing and comparing the covariance matrices of longer and
longer MD trajectories of protein G \cite{Pontiggia2007}, it was seen
that while the trace of the matrix tended to increase (due to the
breadth of visited conformational space), the consistency of the essential
spaces remained highly significant.
{Analogous conclusions were drawn more recently by Liu {\em et al.} who compared the consistency of essential dynamical spaces of cyanovirin-N obtained from atomistic simulations of varying duration\cite{liu_et_al_2011}.}

From these results it emerges that the essential dynamical spaces
calculated from a relatively short MD simulation or from an elastic
network model, would still bear information on the conformational
fluctuations sustained by the proteins over time-scales where the
harmonic approximation is invalid. The fact that these considerations
might hold more in general and not only for the proteins investigated in
refs.\cite{Kitao:1998:Proteins:9849935,Pontiggia2007,Pontiggia:2008:Biophys-J:18931260,liu_et_al_2011}
is reinforced by the fact that the difference vector bridging pairs of
different protein conformers (such as open and closed forms of several
enzymes) has been shown to overlap significantly with the essential
dynamical spaces calculated from elastic network models for either
conformer\cite{Tama:2001:Protein-Eng:11287673,Gerstein.PROTEINS.2002,McCammonAIDS,wall,sanejouand}.

\subsection{Essential dynamical spaces of protein sub-portions}
\label{sec:integration}
For the purpose of comparing the essential dynamical spaces of
proteins with different length and/or architecture it is necessary to
identify the essential dynamical spaces of specific protein subparts.

This is straightforward to do in the context of atomistic molecular dynamics simulations. 
In fact, one simply needs to  restrict considerations to the amino acids of interest when calculating the
average reference structure and the covariance matrix. The top
eigenvectors of this ``reduced'' covariance matrix (whose entries are clearly not equal to the corresponding ones in the matrix computed for the full protein) accordingly
provide the generalised degrees of freedom that best capture the
internal motion of the amino acid of interest.

A different approach is however needed for elastic network models. In this case,
the reduced covariance matrix of the amino acids of interest must be obtained
by the thermodynamic integration of the degrees of freedom of the
remainder amino acids.  For completeness of notation we
assume that the $N$ protein amino acids have been grouped
in two sets, $a$ and $b$. Set $a$ gathers all the $n$ amino acids of interest. 
The interaction matrix $M$, after the row/columns reordering following the amino acid groupings, 
can be partitioned in blocks as follows:

\begin{equation}
M =
\left(
\begin{array}{c|c}
M_a & V   \\
\hline
V^T & M_b \\
\end{array}
\right)
\end{equation}

\noindent where the submatrices
$M_a$ and $M_b$ capture the elastic network interactions involving pairs
of amino acids in set $a$ and $b$, respectively, matrix $V$
contains the elastic network couplings of amino acids in the two sets and $T$ denotes the transpose.
Matrices $M_a$ and $M_b$ are square and symmetric (of linear size $3n$ and $3(N-n)$, respectively) while matrix $V$ is, in general, rectangular.

Because of the quadratic character of the energy function $U$ it is
possible to calculate exactly the reduced matrix effective interactions
for amino acids in set $a$ which is equal to:

\begin{equation}
M^{\rm eff}_a = M_a - V\, M_b^{-1}\, V^T \,
\label{eqn:effM}
\end{equation}
\noindent and finally, the covariance matrix of set $a$ is obtained by
taking the pseudoinverse of $M_a^{\rm eff}$ \cite{hinsen,wall,enzo_JACS}.  It is important to point out that the second term in the right-hand-side of the above equation allows for
taking into account the influence of the remaining amino acids from those
of interest. This term is also crucial to ensure that the dynamics of
amino acids in set $a$ is described in the proper reference system where
the roto-translations of set $a$ alone (and not the whole protein) are
extracted.

We conclude by mentioning that, in the same spirit of
eq.~\ref{eqn:effM}, one can obtain effective interaction (and
covariance) matrices for few generalised degrees of freedom that depend
linearly on amino acid Cartesian coordinates. One such example is
offered by the study of ref.~\cite{Capozzi:2007:J-Proteome-Res:17935310}
where the structural fluctuations of a large set of EF-hand proteins was
studied in terms of the relative motion of the axes of their four
helices.

{A further relevant avenue where the degrees-of-freedom integration can be profitably applied is represented by proteins embedded in a constraining matrix. A notable instance is represented by membrane proteins whose conformational plasticity can have important functional implications\cite{Sachs:2006:Annu-Rev-Biochem:16756508,Engel:2008:Annu-Rev-Biochem:18518819}. For such proteins, Romo and Grossfield \cite{Romo:2011:Proteins:20872850} have recently shown that eq.~\ref{eqn:effM} can be generalised and used to define effective inter-amino acid interactions  which taken into account the influence of embedding bilayer.}

\subsection{Measures of similarities of two sets of essential dynamical spaces}

The information about protein internal dynamics that can be gleaned by
applying the methods described in the previous section, can be used in
quantitative approaches for the dynamics-based comparison, or alignment
of proteins.

We start by discussing the case where the two proteins of interest, $A$
and $B$, are so similar that sequence or structural alignments suffice
to establish extensive one-to-one correspondences between all of their
amino acids or a subset of them.

The consistency of the dynamics of the two sets of amino acids marked
for alignment can be assessed by the standard root mean square inner
product (RMSIP) of their esential dynamical spaces. Customarily, the
comparison is restricted to the top 10 essential modes, which are
usually sufficient to cover most of the global mean square fluctuation
of a protein observed in MD simulations\cite{garcia92}. Accordingly, the
RMSIP is defined as:

\begin{eqnarray}
RMSIP &=& \sqrt{ {1 \over 10} \sum_{l,m = 1}^{10} \left [ \sum_{i = 1}^{n} \sum_\mu  v^l_{i,\mu} w^m_{i,\mu}  \right ]^2}\\
&=& \sqrt{ {1 \over 10} \sum_{l,m = 1}^{10} | {\bf v}^l\cdot {\bf w}^m |^2 }
\label{eqn:rmsip}
\end{eqnarray}

\noindent where ${\bf v}^l$ and ${\bf w}^l$ denote the $l$th essential
mode of the marked amino acids in protein $A$ and $B$, respectively, and
we have further assumed that matching amino acids carry the same index,
$i=1...n$, in the two proteins. Because of the orthonormality of each of
the two basis sets $\{{\bf v}\}$'s and $\{{\bf w}\}$'s, the RMSIP takes
on values in the 0-1 range. 

The RMSIP measure was introduced for the purpose of assessing the
convergence of an MD simulation by comparing the essential dynamical
spaces of e.g. the first and second half of the
trajectory\cite{Amadei99}. Although a simple quantitative criterion for
its statistical significance is lacking, it is generally held that RMSIP
values larger than 0.7 imply meaningful dynamical
similarities\cite{Hess2002}. For completeness we mention that other
measures of dynamical similarity and MD simulation convergence are
available, see e.g. refs.~\cite{Carnevale07,Rueda:2007:Proc-Natl-Acad-Sci-U-S-A:17215349,Fuglebakk:2012:Bioinformatics:22796957,Rueda:2007:Structure:17502102,doi:10.1021/ct050051s}.

We finally point out that, for the purpose of profiling the contribution
of individual amino acids to the overall mean square inner product one
can consider the quantity, which is invariant for changes of the basis
of the essential dynamical spaces\cite{Carnevale07}:

\begin{equation}
Q_i= {1 \over 10} \sum_{l,m = 1}^{10} \left [ \sum_\mu  v^l_{i,\mu} w^m_{i,\mu} \right ]  \left [ {\bf v}^l\cdot {\bf w}^m \right ]
\label{eqn:local_msip}
\end{equation}

\noindent where $i$ is the index of the amino acid of interest, or its square root $q_i=\sqrt{Q_i}$.

\subsection{Best-matching essential dynamical spaces.} 

The RMSIP of eqn.~(\ref{eqn:rmsip}) measures the overall consistency of the essential dynamical spaces
and therefore is invariant upon change of the basis vectors for the two linear spaces, $\{{\bf v}\}$'s and $\{{\bf w}\}$

This property can be exploited to replace the $\{{\bf v}\}$'s and $\{{\bf w}\}$'s
with two new sets of orthonormal vectors  $\tilde{\bf v}^1, \tilde{\bf v}^2,... \tilde{\bf v}^{10}$
and $\tilde{\bf w}^1, \tilde{\bf w}^2,... \tilde{\bf w}^{10}$ which are ranked for decreasing mutual consistency (magnitude of the scalar product)\cite{Pontiggia:2008:Biophys-J:18931260}.

To do so, one constructs the 10x10 asymmetric matrix $D$ whose entries are $D_{ij}={\bf w}_i \cdot {\bf v}_j$.
Next one solves the eigenvalue problems\cite{Pontiggia:2008:Biophys-J:18931260}:
\begin{eqnarray}
D^T\, D {\bf a}^i &=& \mu_i {\bf a}^i\\
D\, D^T {\bf b}^i &=& \mu_i {\bf b}^i\ .
\end{eqnarray}
Assuming that the eigenvalues have been ranked by decreasing order  $\mu_1 > \mu_2 > ... \mu_{10}$ one has that the new basis vectors are given by

\begin{eqnarray}
\tilde{\bf v}^i = \sum_{j=1}^{10} {a}^i_j {\bf v}^j\\ 
\tilde{\bf w}^i = \sum_{j=1}^{10} {b}^i_j {\bf w}^j\\ 
\end{eqnarray}

The newly defined orthonormal basis,  $\{\tilde{\bf v}\}$ and $\{\tilde{\bf w}\}$ have the following remarkable properties:

\begin{itemize}
\item the $i$th vector in one set is orthogonal to all vectors in the {\em other} set with index different from $i$, i.e. 
$\tilde{\bf v}_i \cdot \tilde{\bf w}_j=0$ if $i \not= j$;
\item the scalar products $\tilde{\bf v}_i \cdot \tilde{\bf w}_i$ have magnitude that decreases with $i$
\end{itemize}

\noindent therefore the new basis vectors are optimally ranked for decreasing mutual consistency and are ideally suited to
represent the most consistent (or inconsistent) subspaces spanned by the $\{{\bf v}\}$'s and $\{{\bf w}\}$\cite{Pontiggia:2008:Biophys-J:18931260}.

Once more we stress that, as the  $\{\tilde{\bf v}\}$ and $\{\tilde{\bf w}\}$ provide alternative basis for the same spaces spanned by the  $\{{\bf v}\}$'s and $\{{\bf w}\}$, the RMSIP of  $\{\tilde{\bf v}\}$ and $\{\tilde{\bf w}\}$ is the same as for the  $\{{\bf v}\}$'s and $\{{\bf w}\}$.

\subsection{Beyond structural alignment: dynamics-based protein alignment}

\subsubsection{Aligning proteins by matching their essential dynamical spaces}
\label{sec:aladyn}

The previous approach needs to be suitably generalised in contexts where
one wishes to detect dynamics-based correspondences in different proteins
without relying on their prior sequence or structure alignment.

\begin{figure*}[htp]
\includegraphics[width=0.8\textwidth]{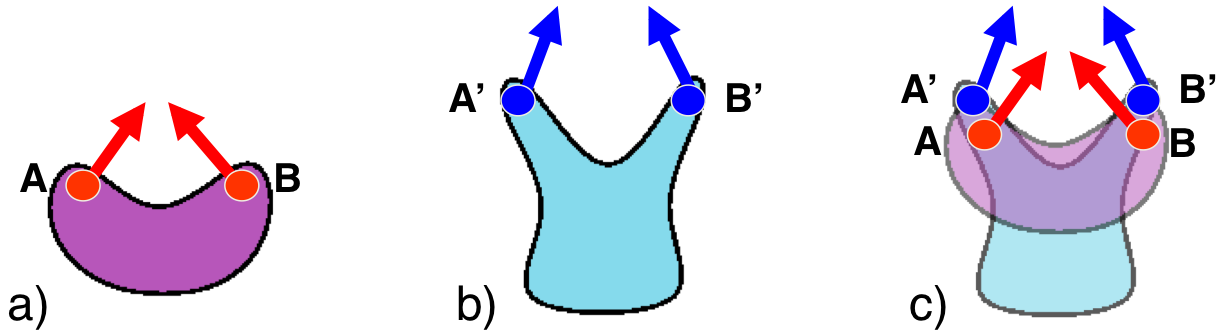}
\caption{Example of dynamics-based alignment. The two cartoon structures
  in panels (a) and (b) have dissimilar shapes. Yet, their internal
  movements, schematically indicated by the arrows, are consistent and can provide
  valuable clues for superposing the two structures, as shown in panel (c).}
\label{fig:cartoon}
\end{figure*}

A prototypical situation is illustrated in Fig.~\ref{fig:cartoon} where
two cartoon structures with different shape are sketched in panels a) and b). 
Despite the overall shape difference, the structural deformation modes described by the arrows, are well-consistent and can provide the basis for aligning the two structures, see panel c).

As first noted by Zen {\em et al.}\cite{Zen:2008:Protein-Sci:18369194},
the example in Fig.~\ref{fig:cartoon} clarifies that meaningful
dynamics-based alignments cannot be simply obtained by purely rewarding
the similarity of directionality and magnitude of the essential
dynamical spaces of any of two sets of amino acids in the proteins of
interest. In fact, the alignment shown in panel c) is intuitively
perceived as viable because the origins of the paired arrows, A--A' and
B--B', are nearby in space. If the origins had been arbitrarily
dislocated in space, then the paired arrows would not have implied any
consistent structural modulations of the two shapes (but motions of very
large amplitude can significantly change the geometrical relationships
of dynamically-corresponding regions, see section \ref{sec:multidom}).

Prompted by the above considerations, Zen {\em et
  al.}~\cite{Zen:2008:Protein-Sci:18369194} introduced and applied a
dynamics-based alignment scheme which simultaneously rewarded the
consistency of the essential dynamical spaces of matching amino acids as
well as their spatial proximity.  Specifically, in this alignment
technique the score to be maximised over the possible sets of
corresponding amino acids pairs was based on distance-weighted
generalization of the root mean square inner product,
\begin{equation}
\sqrt{ {1 \over 10}  \sum_{l,m = 1}^{10} \left [ \sum_{i = 1}^{n} \sum_\mu  v^l_{i,\mu} w^m_{i,\mu} \right ] \, \left [  \sum_{i = 1}^{n} \sum_\mu  v^l_{i,\mu} w^m_{i,\mu} f(d_i) \right ] }
\label{eqn:score}
\end{equation}
\noindent where $i=1,...,n$ runs over the $n$ aligned amino acids, $d_i$ is the distance between the $i$th (matching) amino acids in proteins $A$ and $B$ after an optimal superposition over the putative matching region, and $f(d)=[1-\textrm{tgh}({(d-d_c)/\Delta})]/2$ is a sigmoidal distance weighting factor where $d_c=4$\AA\ and $\Delta=2$\AA.  

Notice that, as for the RMSIP, the measure (\ref{eqn:score}) is independent of the choice of the bases spanning the linear space of the top 10 essential dynamical modes. 

The sought dynamics-based alignment is accordingly obtained by
maximizing the measure of eq.~\ref{eqn:score} (after a suitable
$n$-dependent regularization, see
ref.~\cite{Zen:2008:Protein-Sci:18369194}) over the space of possible
amino acid pairings in the two proteins, and finally by assessing its
statistical significance by comparing it against a null reference case.

Clearly, the combinatorial space of matching amino acids is very large and,
because each attempted alignment involves the re-calculation of the
essential dynamical spaces, the computational effort entailed by this
comparison is significant and can take several minutes on present-day computers
for two proteins of $\sim 100-200$ amino acids.

By heuristically restricting the search of matching amino acids and by using approximate but faster calculations of the alignment score, the original algorithm of Zen {\em et al.} ~\cite{Zen:2008:Protein-Sci:18369194} was sped up sufficiently for interactive use via the Aladyn web-server\cite{Potestio:2010:Nucleic-Acids-Res:20444876}. The results of this publicly-available server will be frequently referred to in the remainder of this article.

\subsubsection{Aligning proteins by matching pairwise distance fluctuations}
\label{sec:biggin}

An alternative method to align proteins based on their internal dynamics
properties was recently proposed by Biggin and coworkers\cite{Munz:2010:BMC-Bioinformatics:20398246}. In this method one exclusively considers the pairwise distance
fluctuations of amino acids, with no explicit reference to the spatial
coordinates of the latter, nor to the detailed information contained in the top essential dynamical spaces. This scheme is based on the idea that, if a set of
amino acids $\{\alpha\}$ in protein A has similar movements to a
corresponding set of amino acids $\{\beta\}$ in protein B then the
matrices of pairwise distance fluctuations of the two sets, $F_\alpha$
and $F_\beta$, should be similar too.

In the approach of  M{\"u}nz {\em et al.}~\cite{Munz:2010:BMC-Bioinformatics:20398246} a generic entry of the $F$ matrix is defined as
\begin{equation}
F_\alpha(i,j)= {\rm std.dev}(d_{\alpha_i,\alpha_j})
\end{equation}
\noindent where the right-hand-side is the standard deviation of the
distance of amino acids $i$ and $j$ in set $\alpha$ calculated over a
converged molecular dynamics trajectory.

Next, one calculates the relative difference of each corresponding matrix entry,
\begin{equation}
d(i,j)= {|F_\alpha(i,j)-F_\beta(i,j)| \over \left ( F_\alpha(i,j)+F_\beta(i,j) \right )/2}
\label{eqn:dist}
\end{equation}
\noindent  
\noindent and an overall dynamical score $S^{AB}(\alpha,\beta)$ is constructed by
weighting the contribution of all $d(i,j)$'s.

As in the previous approach, the best dynamics-based alignment of the
two proteins is found by maximising $S^{AB}(\alpha,\beta)$ (again after
a suitable length-regularization procedure) over all possible choices of
$\{\alpha\}$ and $\{\beta\}$. In the study of
ref.~\cite{Munz:2010:BMC-Bioinformatics:20398246}, the exploration of
the vast combinatorial space of a.a. pairings was carried out within a
Monte Carlo optimization scheme.

\subsubsection{Aligning proteins by matching the mean square fluctuation profiles}
\label{sec:tobi}

The possibility to align proteins by detecting correspondences in the
amplitudes of amino acids motions in different proteins was first
explored by Keskin {\em et al.}  \cite{keskin}. In this study, which is
covered in section \ref{sec:rossmann}, the one-to-one correspondences of
amino acids in a set of structurally-related proteins was based on a
supervised matching of the amplitude of amino acid fluctuations computed
from an isotropic elastic network model\cite{bahar97}.

An automatic implementation of this alignment strategy was recently introduced by
Tobi ref.~\cite{Tobi:2012:Proteins:22275069}. In this study, the
one-dimensional character of the quantity to be matched (mean square
fluctuation) was exploited, as in sequence alignments, within the
dynamical-programming alignment of Needleman and
Wunsch\cite{Dynamicprogramming}.

\section{Comparative studies of protein internal dynamics}

Early systematic dynamics-based comparisons were all targeted to groups
of proteins known to be significantly related from the sequence,
structural or functional point of view. In such contexts, in fact, the
assessment and interpretation of the comparisons is more
straightforward.  Accordingly, we shall first discuss these comparative
investigations of proteins whose relatedness is known {\em a priori}. We
shall next report on studies which considered proteins with limited
structural relatedness as well as investigations targeted at
understanding more general (and possibly evolutionary) dynamics-based
aspects of the structure/function relationship. When appropriate, the
results of these earlier studies will be revisited using the
dynamics-based alignment of ref.~\cite{Zen:2008:Protein-Sci:18369194}
as implemented in the publicly available Aladyn web-server[75].

\subsection{Common fluctuation patterns in proteins with a Rossmann-like fold}
\label{sec:rossmann}

We first discuss the case of proteins adopting a Rossmann-like fold
which were addresses in the studies of Keskin {\em et al.}\cite{keskin}
and Pang {\em et al.}\cite{sansom}.

In the study of ref.~\cite{keskin}, which is arguably the first
dynamics-based comparative investigation, Keskin {\em et al.} considered
six proteins each consisting of two linked globular domains with a
Rossmann-like fold.  The proteins covered two homologous groups: the
first one (CATH\cite{CATH} code 3.40.190.10) included cofactor binding
fragment of CysB, the lysine/arginine/ornithine-binding protein (LAO),
the enzyme porphobilinogen deaminase (PBGD), the N-terminal lobe of
ovotransferrin (OVOT) while the second one (CATH code 3.40.50.2300)
comprised the ribose-binding protein (RBP) and the
leucine/isoleucine/valine-binding protein (LIVBP).

The internal dynamics of these proteins was characterised by using a
simplified (isotropic) Gaussian network model \cite{bahar97} to compute
their mean-square fluctuation profiles and the lowest energy modes. The
authors observed that the latter mostly entailed a hinge-bending motion
of the two domains around the linker and the predicted motion amplitude
varied significantly between the unliganded and liganded state of the
molecules. In connection to this latter result it is worth noting that
for several other proteins it has been shown that the internal dynamics
sensitively depends on substrates and cofactors. A prototypical example
is offered by dihydrofolate reductase where dynamical properties,
arguably linked to catalysis, has been shown numerically to strongly
depend on the type of bound ligand\cite{Radkiewicz:2001:J-Am-Chem-Soc:11472122}.

The similarities of the modes amplitude profiles across the six
proteins, further prompted Keskin {\em et al.} to attempt a
manually-curated alignment of the proteins by matching the modes shape
in a gapless portion of one of the two domains. The amino acid
correspondences were next extended to the remainder of the proteins by
inspecting both their FSSP structural alignments
\cite{Holm:1994:Nucleic-Acids-Res:7937067} and, again, the modes shape.
These supervised alignments returned very good superpositions of the
modes amplitude profiles across the considered proteins and, because of
the limited use of structural correspondences, the RMSD after an optimal
superposition of the corresponding amino acids was about 7\AA.

From the consistency of the modes' profiles the authors concluded that
members of the same fold can share common dynamical features on a
global, collective scale and further envisaged that fully-automated
dynamics-based alignments of proteins might have been feasible.

The implications of structural relatedness for the similarity of protein
internal dynamics were next explored by Pang {\em et al.}\cite{sansom}
by using atomistic molecular dynamics simulations on a set of four
periplasmic binding proteins in various forms: apo, holo and
crystallized in different conditions. 

The monomeric units of these entries, which included the LAO protein
considered by Keskin {\em et al.}\cite{keskin}, comprised about 230
amino acids and consisted, again, of two Rossmann-like domains connected
by a linker. Based on DALI\cite{Holm:2000:Bioinformatics:10980157} alignments Pang
{\em et al.} identified a core of 100 amino acids (i.e. spanning
about 40-45\% of the proteins) common to the four proteins.

The comparison of the internal dynamics was carried out on the common
core amino acids and regarded various quantities calculated from 10- or
20-ns long molecular dynamics simulations.  In particular, the
comparison included: the amino acids' mean square fluctuations, the
overlap of the covariance matrices and the overlap (RMSIP) of the two
essential dynamical spaces.

By comparing the properties of the same protein but in liganded and
unliganded forms, Pang {\em et al.}  observed clear differences in the
molecules' internal dynamics, consistently with the findings of Keskin
{\em et al.} reported above.

 Regarding the comparison of different proteins, the
 authors reported a significant overlap of all dynamical properties
 computed over the common core.  In particular, throughout the set of
 periplasmatic binding proteins, the first and second essential
 dynamical modes systematically corresponded to, respectively, the
 hinge-bending and twisting motions of the linked domains.

However, by examining how the overlap of the covariance matrix and
essential dynamical spaces increased with simulation time, the authors
observed that each protein tended to occupy specific regions of the
essential dynamical space. It was concluded that these differences
reflected protein-specific features, arguably encoded in their sequence. While, it cannot be ruled out {\em a priori} that the the
observed differences could be ascribable to the several non-aligned
amino acids, the observation of Pang {\em et al.} is very interesting
and relevant in the present context, because it points to specific
dynamics-based features which can be beyond reach of
sequence-independent approaches, such as elastic network models.

\subsection{Dynamics-based alignment of proteases}

\begin{table}
\begin{tabular}{|l|l|l|l|}
  \hline
tag &  protein &  PDBid &  CATH code \\
  \hline \hline
A     & endothiapepsin (ASP)           & 1er8E  & 2.40.70.10\\
B     & HIV-1 protease (ASP)        & 1nh0AB & 2.40.70.10\\
C     &  3C-like proteinase (SER)      & 1uk4A  & 2.40.10.10 \\
& & & 1.10.1840.10 \\
D$_1$ & adenain (CYS)                  & 1avp   & 3.40.395.10\\
D$_2$ &sedolisin (SER)                 & 1ga6   & 3.40.50.200 \\
D$_3$ &pyroglutamyl peptidase I (CYS)  & 1ioi   & 3.40.630.20\\
E     & assemblin (SER)                & 1jq7A  & 3.20.16.10\\
F$_1$ &dipeptidyl-peptidase I  (CYS)   & 1k3bA  & 2.40.128.80\\
F$_2$ &cruzipain                       & 1me4   & 3.90.70.10\\
G$_1$ &atrolysin E (Zn)                & 1kuf   & 3.40.390.10\\
G$_2$ & carboxy peptidase A1 (Zn),     & 8cpa   & 3.40.630.10\\
  \hline
\end{tabular}
\caption{Representatives of the seven common protease folds, A--G. The
  list includes proteases with different catalytic chemistry (aspartic-,
  serine-, cystein- and metallo-proteases). For convenience of
  comparative purposes, because the active site is comprised within the
  monomeric units of 3C-like proteinase, assemblin and
  dipeptidyl-peptidase I, we did not consider the multimeric biological
  form of these entries. Conversely, because the catalytic aspartic dyad
  of HIV-1 protease straddles the dimeric interface, we retained its
  full dimer. The corresponding structures are represented in Fig.~\ref{fig:proteases}.}
\label{tab:proteases}
\end{table}

Proteases, enzymes that cleave peptide chains, account for about
2\% of the genome of various organisms\cite{Rawlings:2012:Nucleic-Acids-Res:22086950,Quesada:2009:Nucleic-Acids-Res:18776217,Southan:2000:J-Pept-Sci:11016882}. In view of this representative weight
and biological importance, they have been systematically investigated
and compared.

\begin{figure*}[htp]
\includegraphics[width=0.95\textwidth]{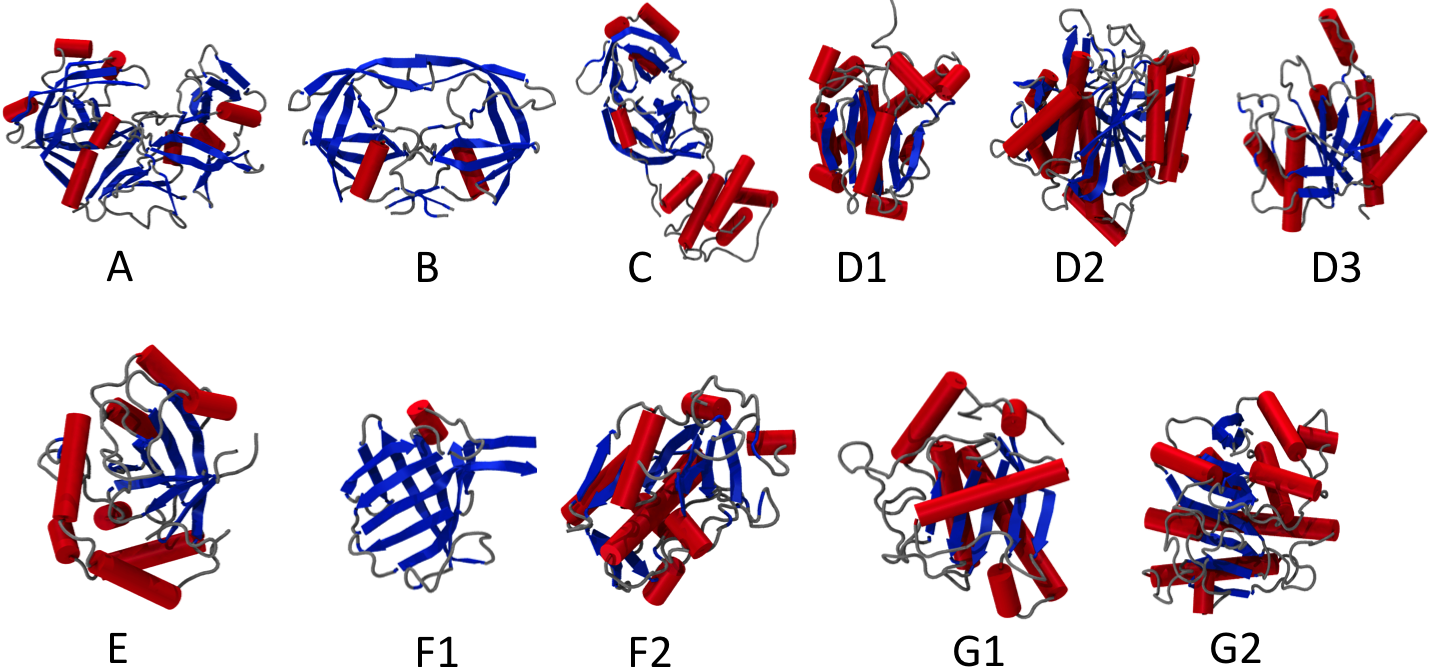}
\caption{Representative structures of the common protease folds
  listed in Table \ref{tab:proteases}. This illustration and subsequent ones
  were prepared with the VMD graphical package\cite{VMD}.}
\label{fig:proteases}
\end{figure*}

The comprehensive survey carried out by Tyndall {\em et
  al.}\cite{tyndall}, identified 7 common structural folds for this
family of enzymes. Various representatives for the seven common folds
were identified by Carnevale {\em et al.}\cite{enzo_JACS} and are listed
in Table~\ref{tab:proteases} and shown in Fig.~\ref{fig:proteases}.

As reported in Table \ref{tab:proteases}, the various representatives
cover 4 different architectures and 9 different topologies of the CATH
classification scheme\cite{CATH}. Notice that the two aspartic
proteases, the endothiapepsin and HIV-1 PR share the full CATH code,
implying that they have detectable sequence homology despite their their
marginal sequence identity, different length and different oligomeric
state (monomeric for edothiapspsin and dimeric for HIV-1 PR)\cite{tang,blundell,cascella05}.

Besides this ASP-protease pair, other pairs of entries listed in Table
\ref{tab:proteases} have significant overall structural similarities. In
particular the six possible distinct pairings between pyroglutamyl
peptidase I, atrolysin E, sedolisin and carboxy peptidase A1 are all
significant according to the DALI statistical
criteria\cite{Holm:2000:Bioinformatics:10980157}.  Interestingly, the
simultaneous multiple alignment of these four entries is poor and
involves several short fragments for a total of about 30 amino acids
(consistently for both Mistral and Multiprot\cite{Mistral,Multiprot}).

\begin{figure*}[htp]
\includegraphics[width=0.9\textwidth]{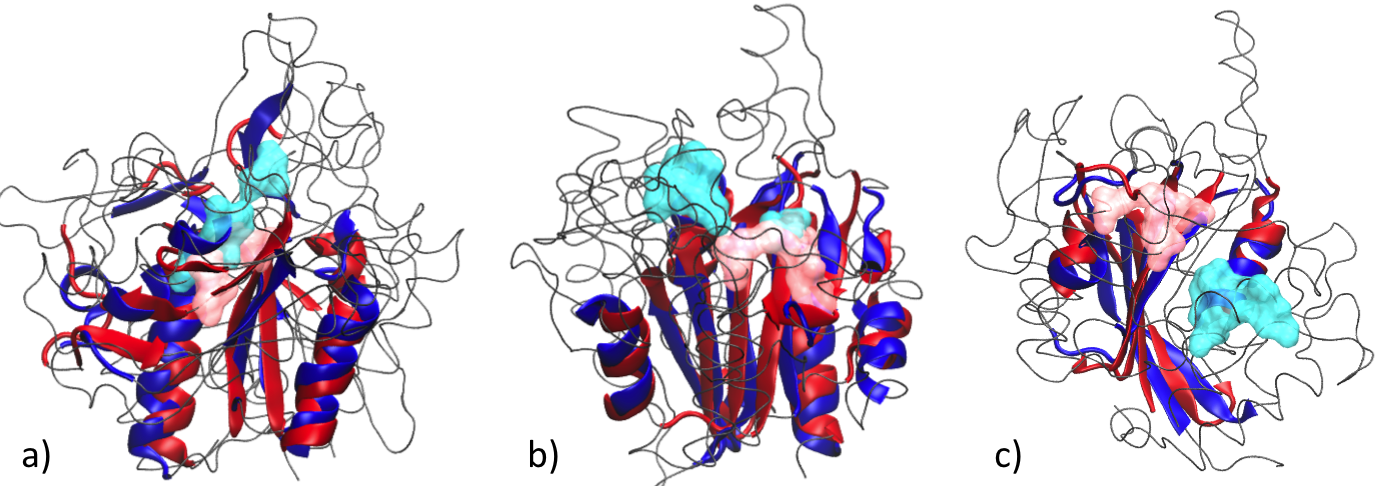}
\caption{Stuctural alignments of pyroglutamyl peptidase I with a)
  sedolisin, b) carboxy peptidase A1 and c) atrolysin E. In all panels
  the pyroglutamyl peptidase I is shown in red, while the partner
  proteins are shown in blue.  Aligned regions are shown with thick
  ribbons and known active sites\cite{CSA1} are highlighted with
  Van-der-Waals surfaces. The trace of non-aligned regions is shown as a thin grey curve.}  
\label{fig:1ioi}
\end{figure*}

The top structural alignments within this group involved the entry
pyroglutamyl peptidase I and are shown in Fig.~\ref{fig:1ioi}.  As it
  was reported in ref.~\cite{enzo_JACS} (see Fig. 3 therein) the
  alignments, involve several disconnected matching fragments comprising
  the active site and the surrounding region within 7-10 \AA\ of it.

The good structural superposition of the active sites in panels (a) and
(b) of the Figure provides evidence for the existence of
functionally-related traits that are shared by proteases that are
non-homologous and rely on different catalytic chemistry (serine,
cysteine- and metallo-proteases).
 
\begin{figure*}[htp]
\includegraphics[width=0.9\textwidth]{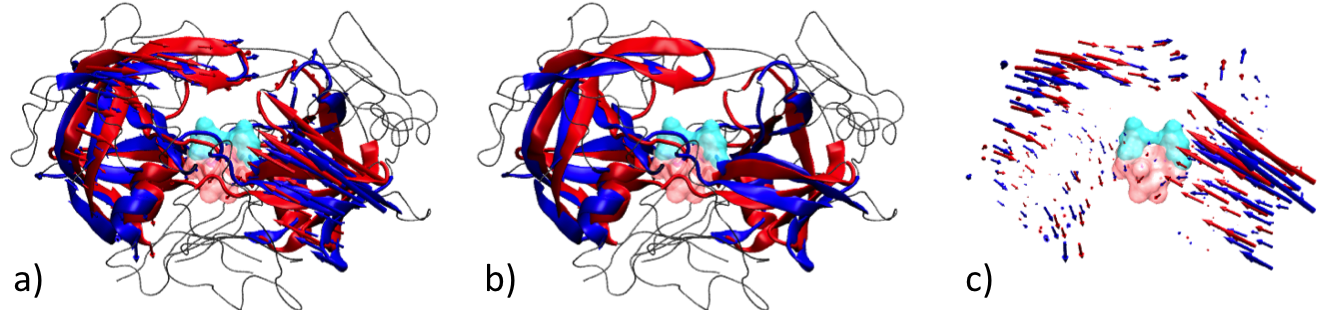}
\caption{(a)Dynamics-based alignment of HIV-1 protease (red) and
  endothiapespsin (blue) obtained with the Aladyn
  web-server\cite{Potestio:2010:Nucleic-Acids-Res:20444876}. A thick
  ribbon is used to highlight aligned regions and known active sites are
  highlighted with Van-der-Waals surfaces. The trace of non-aligned
  regions is shown as a thin grey curve while the arrows represent the
  three best matching essential modes. The ribbons and the modes are
  shown separately in panels (b) and (c), respectively.}
\label{fig:hiv_bace_dyn}
\end{figure*}

The fact that functional activity of various proteases is known to be
impacted by their large-scale internal
dynamics\cite{blundell,piana02a,piana02b,McCammonAIDS}, which can
involve mechanical couplings between the active site and distal regions
at the protein surface \cite{piana02a,micheletti04,piana02b,McCammonAIDS}, poses the question of whether
dynamics-based alignments can be used to identify further relationships
between proteases that are elusive to the pure structural comparison.
The possibility to do so is illustrated in Fig.~\ref{fig:hiv_bace_dyn},
which illustrates the dynamics based alignment of HIV-1 PR and
endothiapepsin.

\begin{figure*}[htp]
\includegraphics[width=0.95\textwidth]{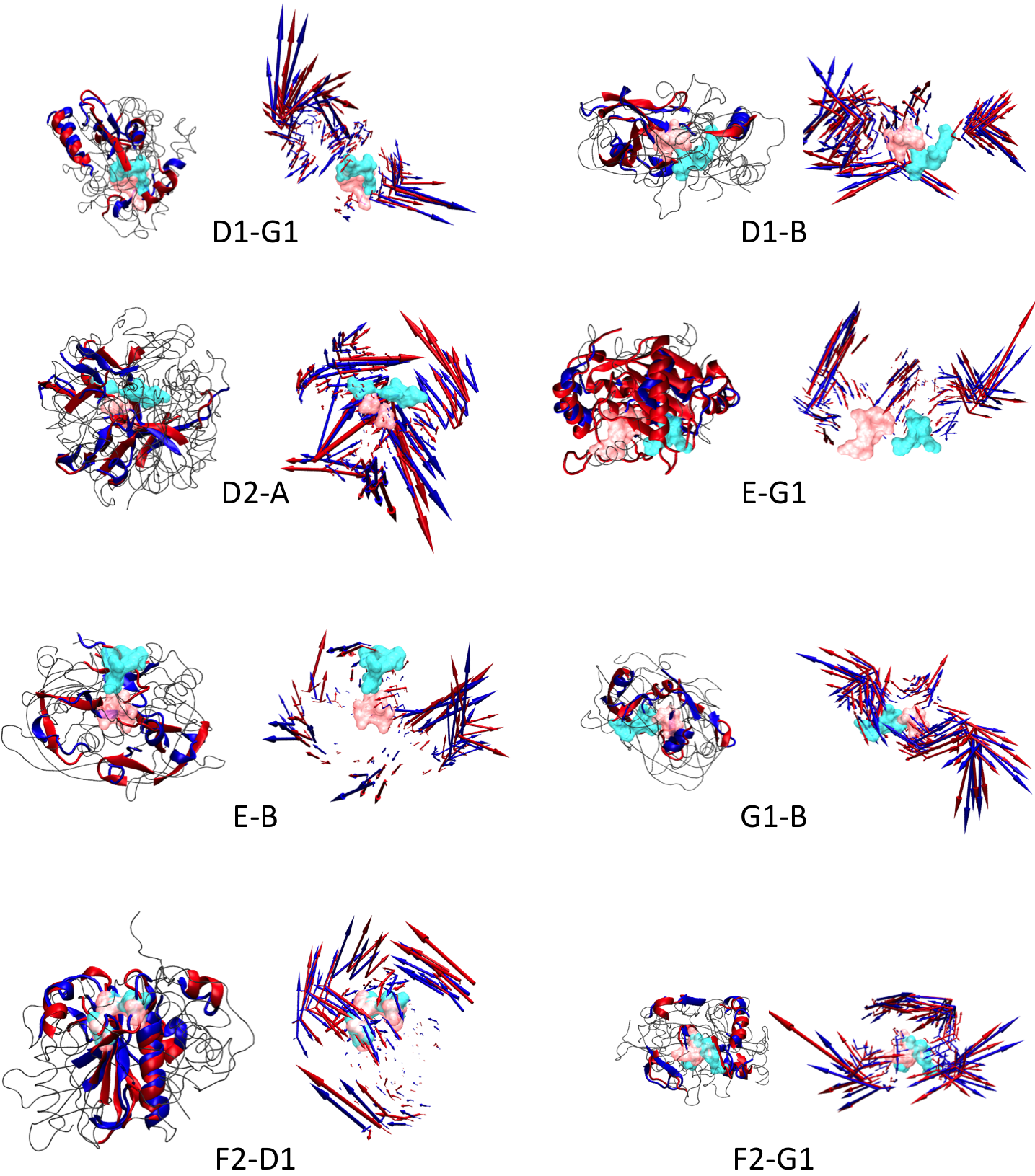}
\caption{Significant dynamics-based alignments of various pairs of proteases. The pairs are tagged as in Table \ref{tab:proteases}.
For each pair we report separately the structural superposition of the aligned regions (ribbons) and of the top three best-matching modes (arrows). Aligned elements are shown in blue for the first entry of the pair and in red for the second. The active sites are shown in cyan and pink for the first and second entry of the pair, respectively.}
\label{fig:composita_dyn}
\end{figure*}

Following the spirit of ref.~\cite{enzo_JACS}, we have used the Aladyn
algorithm to align all pairs of entries in Table~\ref{tab:proteases}.
In addition to the previously mentioned significant structural pairings,
the Aladyn algorithm identifies 8 additional significant alignments
($p$-value $<0.02$, corresponding to the incidence of less than one
false positive in the set of all pairwise alignments of the entries in
Table~\ref{tab:proteases}).  These pairs are shown in
Fig.~\ref{fig:composita_dyn}.  Notice that calpain, adenain, atrolysin E, and HIV-1PR (corresponding respectively to tags F$_2$,  D$_1$, G$_1$ and B in Table \ref{tab:proteases}) constitute a notable dynamically-alignable
``clique'' because all pairings of these proteins (with the sole
exception of cruzipain--HIV-1PR which involves only 30 amino acids) are
significant.

The structural and dynamical consistency of the 8 aligned pairs is shown
in Fig.~\ref{fig:composita_dyn}. 
It is striking to see that the active sites of the compared proteins are
very well superposed or in contact, with the exception of two
alignments, assemblin--HIV-1 PR (E--B) and assemblin-- atrolysin E
(E--G$_1$) where the active sites are at a distance of 10\AA.  The
overall RMSD of the matching amino acids is $\sim 3.0$\AA.

It is also noticed that the corresponding modes, tend to outline a
shearing deformation of region surounding the active site.  This result
is in accord with the general functional features common to proteases,
which consists of the shearing of the bound peptide into a beta extended
conformation prior to cleavage\cite{tyndall}. More generally, the
finding is consistent with the observed property that active sites in
enzymes tend to be located at the interface of quasi-rigid domains, as
this can ensure a fairly rigid geometry of the catalytic region located
at the interface combined with an appreciable modulation of the
surrounding region which ought to aid the substrate recognition and
processing
\cite{SacquinMora:2007:Proteins:17311346,Potestio:2009:Biophys-J:19527659}.

For the specific case of proteases, the dependence of the enzymatic
activity and catalytic rate on the global conformational fluctuations of
the proteins has been advocated for HIV-1 PR\cite{piana02b} (but this
does not occur for furin, a serine protease\cite{Carnevale}). The proposed mechanism for HIV-1 PR has been corroborated by recently experimental findings\cite{Das:2010:J-Am-Chem-Soc:20397633}.
Further examples of the coupling between the modulation of the geometry of the
region near the active site and the global protein motions are provided
by triose phosphate isomerase\cite{Kurkcuoglu:2006:Biochemistry:16430213}
and dihydrofolate reductase\cite{Agarwal:2002:Proc-Natl-Acad-Sci-U-S-A:11867722,rod}

We emphasize that all the pairings identified with the dynamics based
alignment shown in Fig.~\ref{fig:composita_dyn} are not deemed
significant in DALI alignments. The findings therefore suggest that, for
certains proteins and enzymes, some functionally-oriented features can
be more confidently identified using dynamics-based alignments than with
sequence- or stucture-based alignment approaches.

\subsection{Dynamics-based alignment of PDZ domains}

\begin{table*}
\begin{tabular}{|l|l|l|l|}
  \hline
Compound &  CATH domain &  CATH code \\
  \hline \hline
 Postsynaptic density protein 95  (PSD-95) & 1be9A00 & 2.30.42.10\\
nitric-oxide synthase  (nNOS)& 1qauA00 & 2.30.42.10 \\
 Alpha-1 syntrophin  & 1qavA00 & 1.14.13.39\\
Inactivation-no-after-potential D protein (Inad)  & 1ihjA00 & 2.30.42.10 \\
Segment polarity protein dishevelled homolog DVL-2 (DVL2)  & 2f0aA00 & \\
Glutamate receptor interacting protein 2 (GRIP2) & 1x5rA00 & \\
\hline
Tricorn protease  & 1k32A04 &  \\
 	
Type II secretion system protein C & 2i6vA00 & 3.4.21\\
 hypothetical serine protease rv0983  & 1y8tA03 \\
 Photosystem II D1 protease & 1fc6A02 & 2.30.42.10\\
  \hline
\end{tabular}
\caption{List of PDZ domains considered in ref.~\cite{Munz:2010:BMC-Bioinformatics:20398246}. The line serapates PDZ domain from multicellular organisms (above line) from  unicellular ones (below line).}
\label{tab:PDZ}
\end{table*}

We next discuss the dynamical similarities of members of the PDZ domain
family. PDZ domains are structural moduli commonly associated to ion
channels and receptors or otherwise involved in signal transduction
pathways\cite{Songyang:1997:Science:8974395,Fanning:1999:J-Clin-Invest:10079096,Munz:2010:BMC-Bioinformatics:20398246,Biggin2012}.
\begin{figure*}[htp]
\includegraphics[width=0.95\textwidth]{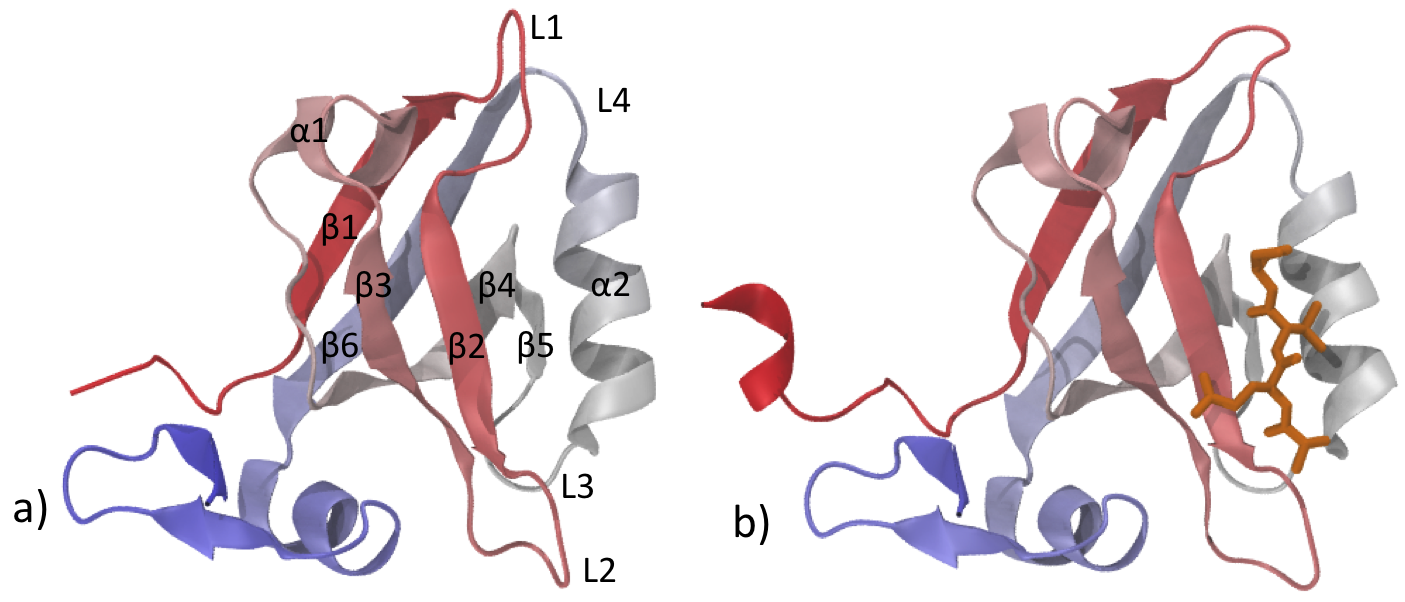}
\caption{(a) Apo and (b) holo forms of a PDZ domain. The PDBid of the shown entries is 1bfe for the apo form and 1be9 for the holo one. The ligand bound to the $\alpha_2$-$\beta_2$ cleft of the holo form is highlighted in orange.}
\label{fig:PDZ}
\end{figure*}

They are typically 80-100 amino acids long and adopt an overall globular
fold comprising two $\alpha$ helices and 6 $\beta$ strands, see
Fig.~\ref{fig:PDZ}b. The interaction with a partner protein usually
occurs through the accommodation of its C-terminal segment in the
$\beta_2$-$\alpha_2$ cleft.  In fact, the observed mobility of helix
$\alpha_2$ relative to the PDZ-domain core has been argued to be
important for ligand binding and recognition
\cite{Munz:2010:BMC-Bioinformatics:20398246,delos,Kong:2009:Proteins:18618698}. Although
PDZ-domains sustain modest structural changes after ligand binding, see
panels a and b in Fig.~\ref{fig:PDZ}, experimental and numerical
evidence suggest that there exist allosteric pathways running internally
to the molecule that signal the binding event to regions that are
opposite on the protein surface respect to the binding cleft
\cite{ranganathan99,Kong:2009:Proteins:18618698,delos,Law:2009:J-Am-Chem-Soc:19374353}. While
key aspects of the signal propagation mechanism are still controversial
\cite{Chi:2008:Proc-Natl-Acad-Sci-U-S-A:18339805} various evolutionary
aspects of the allosteric mechanism and the binding mode have been
actively investigated {using a variety of techniques including bioinformatics \cite{ranganathan99},  NMR \cite{Law:2009:J-Am-Chem-Soc:19374353}, elastic network linear response theory \cite{gerek_ozkan_PDZ} and molecular dynamics simulations \cite{Munz:2010:BMC-Bioinformatics:20398246}.}

In particular, Biggin and coworkers
\cite{Munz:2010:BMC-Bioinformatics:20398246} have recently introduced
and systematically applied the dynamics-based alignment outlined in
section \ref{sec:biggin}, to compare the mainchain dynamics of 10 PDZ
domains from both unicellular and multicellular organisms, see
Table~\ref{tab:PDZ}.  The dynamics-based comparison, was based on the
analysis of pairwise distance fluctuations of amino acids calculated
from 20ns-long atomistic molecular dynamics simulations.

Within this set of sequence- and structurally-related PDZ domains Munz
{\em et al.} observed the largest dynamical consistency among the
domains from multicellular organisms.  In fact, significant
dynamics-based similarities were found almost exclusively among entries
from multicellular organisms (particularly pairs nNOS--PSD95,
nNOS--alpha-1 syntrophin, nNOS--DVL2, Inad--Alpha-1 syntrophin,
Inad--DVL2, DVL2--Alpha-1 syntrophin).

\begin{figure*}[htp]
\includegraphics[width=0.95\textwidth]{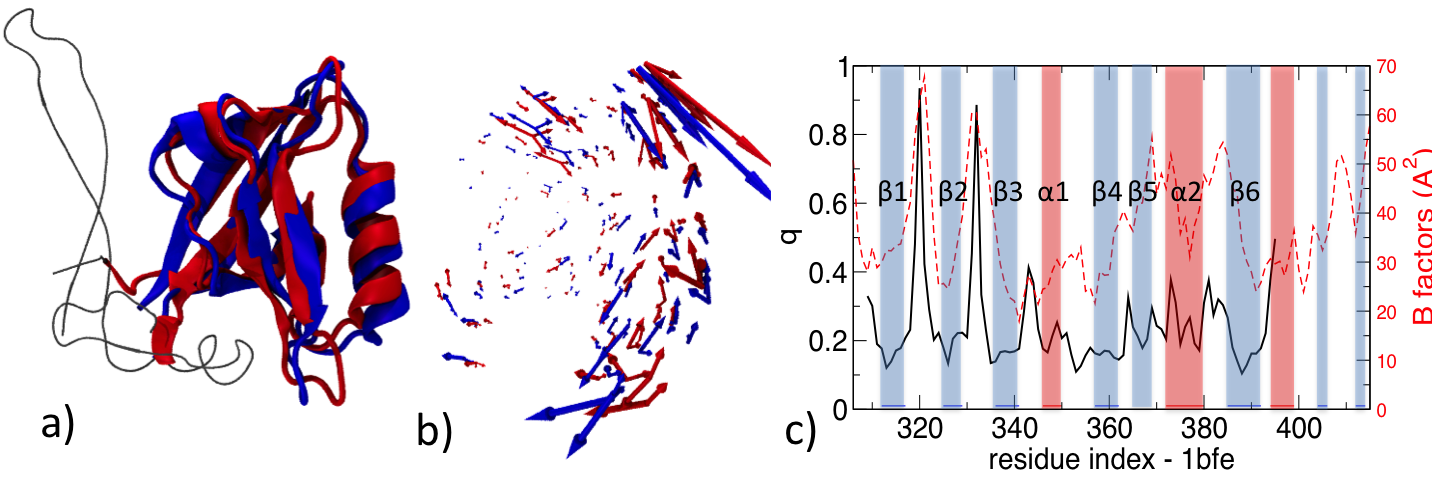}
\caption{Dynamics-based alignment \cite{Potestio:2010:Nucleic-Acids-Res:20444876} of the two PDZ domains discussed in 
ref.~\cite{Munz:2010:BMC-Bioinformatics:20398246}. 
The alignment was obtained with the  obtained with the Aladyn
  web-server\cite{Potestio:2010:Nucleic-Acids-Res:20444876} and consists of an uninterrupted stretch 87 amino acids (ARG309-GLU395 for 1bfe and ASN14-GLU101 for
1qau) at an RMSD of 2.2\AA\ and with an RMSIP of 0.74. The structural superposition is shown in panel (a) and the top three matching modes are shown in panel (b). Corresponding elements for entry 1bfe are shown in red while those for entry 1qau are shown in blue.
The crystallographic B--factors and the local essential dynamics space overlap, $q=\sqrt{Q}$, (see eqn. \ref{eqn:local_msip}) of 1bfe are shown respectively with a dashed and a solid line in panel c.}
\label{fig:PDZ_alignments}
\end{figure*}

One such pair, PSD95 and nNOS, was analysed in-depth to highlight the
differences of sequence, structure and dynamics-based
alignments. Through this comparative investigation, the authors noticed
that dynamical correspondences were particularly poor in the $\alpha_2$
region, which is otherwise structurally well-alignable. 
Because the mobility of this helix arguably impacts the binding
of ligands it was concluded that the dynamical differences could reflect
subtle differences in the functionality of PSD95 and nNOS
\cite{Munz:2010:BMC-Bioinformatics:20398246}.

The findings of Munz {\em et al.}, are illustrated and revisited here
through the dynamics-based alignment method of Zen {\em et al.} as implemented
in the Aladyn web-server. The Aladyn alignment of PSD95 and nNOS is
shown in Fig.~\ref{fig:PDZ_alignments} and illustrates the good
consistency of the essential dynamical spaces of the aligned regions.
Interestingly, the contribution of the various corresponding amino acids to the good
RMSIP value, which is equal to 0.74, is rather uneven.

This is illustrated in Fig.~\ref{fig:PDZ_alignments}b which portrays the
residue-wise contribution to the mean square inner product, $Q_i$ (see
eq.~\ref{eqn:local_msip}) along with the mean-square residue
fluctuations.  It is seen that the $Q$ profile is peaked in
correspondence of the loops $L_1$, $L_2$, $L_3$ and $L_4$ which are also
associated to peaks of the crystallographic B-factor profiles.
Although the comparison of computed mean-square fluctuations with B-factors is not perfectly transparent (the latter are affected by crystal packing and disorder \cite{Frauenfelder:1988:Annu-Rev-Biophys-Biophys-Chem:3293595}), the
accord of the two sets of peaks is consistent with the intuition that,
given the overall accord of the essential modes, the highest values of
$Q$ should be observed in correspondence with regions of high mobility
(where the norm of the essential modes concentrates).  By the same
token, one would have expected to observe a peak of the $Q$ profile in
correspondence of the mobile helix $\alpha_2$ and the nearby portions of
the flanking strands $\beta_5$ and $\beta_6$. By contrast, however, the
relative contribution of these regions to the RMSIP is small. This is
therefore indicative of a poor consistency of the generalised direction
of motion of this region in the two proteins of interest, thus
confirming the findings of Munz {\em et al.} from a different
dynamics-based perspective.

\subsection{Conservation of general dynamical patterns in protein families and superfamilies}

Besides the previous investigations that aimed at elucidating specific
functionally-related aspects in different proteins by using
dynamics-based alignment strategies, there have been a number of studies
where more general dynamical properties were compared across various
protein families and superfamilies.

In recent years Echave and coworkers have carried out several such
studies with the purpose of assessing the extent to which features such
as mean-square fluctuation profiles and overall shape (amplitude modulation) of the
essential modes have been evolutionarily conserved
\cite{Maguid:2005:Biophys-J:15749782,Maguid:2006:J-Mol-Evol:17021932,Maguid:2008:Gene:18577430}.

The first of such analyses was carried out for a set of 18 members of
the globin family\cite{Maguid:2005:Biophys-J:15749782}.  The considered
globins typically consisted of 130-150 amino acid and shared a
structural core of 68 amino acids \cite{Mistral}.

The comparison of Maguid {\em et al.}
\cite{Maguid:2005:Biophys-J:15749782} was focused on the set of about
100 corresponding amino acids that were identified by the multiple
(CLUSTAL \cite{Thompson:1997:Nucleic-Acids-Res:9396791}) sequence
alignment of the 18 globins.

The dynamics of the globins was next characterised by the mean-square
fluctuation profiles and molecules' lowest energy modes which were
computed using the isotropic Gaussian network model \cite{bahar97}. In this model,
the presence of the heme group was not taken into account.
 
The comparison of the the dynamics across the different globins was
carried out by measuring the linear correlation coefficient between the
fluctuation amplitudes of corresponding amino acids or between their
displacements in the top modes. For comparative purposes, the latter
were reranked so as to have maximally compatible sets of first modes,
second modes etc. across the globins. The main differences of this
comparative strategy from the one described in section
\ref{sec:integration} is that the dynamics of the corresponding amino
acids is obtained by neglecting the effect of non-aligned amino acids
(equivalent to omitting the second term in eq.~\ref{eqn:effM}) and for
the use of reranked top modes in place of identifying the most
consistent directions in the linear space spanned by the top modes.

After carrying out these comparative steps, Maguid {\em et al.}
\cite{Maguid:2005:Biophys-J:15749782} concluded that both the
mean-square fluctuations and the shape (amplitude modulation) of the top
reranked modes were highly consistent across the various members of the
globin family.

Building on this findings, Maguid {\em et al.}
\cite{Maguid:2006:J-Mol-Evol:17021932,Maguid:2008:Gene:18577430}
extended the analysis to a a comprehensive set of $\sim 1000$ protein
entries from several hundred families superfamilies of the HOMSTRAD
database\cite{Mizuguchi:1998:Protein-Sci:9828015,Stebbings:2004:Nucleic-Acids-Res:14681395,Pugalenthi:2005:Nucleic-Acids-Res:15608190}.
The studies followed the same comparative pathway outlined above for the
globins, with the significant modifications that corresponding amino
acids were identified in pairwise MAMMOTH
\cite{Ortiz:2002:Protein-Sci:12381844} structural alignments and the
anisotropic beta-Gaussian {elastic} network model was used in place of the
isotropic one. Furthermore, the degree of collectivity of the modes was
also assessed and compared.

The studies of
refs.~\cite{Maguid:2006:J-Mol-Evol:17021932,Maguid:2008:Gene:18577430}
reported that the dynamical similarity (mean-square-fluctuation
profiles, mode shape and mode collectivity) within members of the same
family and superfamily was significantly larger compared to pairs of
unrelated protein entries. In addition, the similarity within the same
family was stronger than within the same superfamily.

From this series of studies, Echave and coworkers concluded that general
dynamical properties of proteins tend to be preserved in the course of
evolution and are quantitatively detectable.

\subsection{Conservation of specific functionally-oriented dynamics in enzymes}

In the recent study of ref.~\cite{Ramanathan:2011:PLoS-Biol:22087074}
Ramanathan {\em et al.}  addressed, by means of atomistic molecular
dynamics simulations, the extent to which enzymes with the same function
but different degree of homology rely on the same functionally-oriented
dynamics.

The study considered a few members for each of three different types of
enzymes: the CypA peptidyl-prolyl isomerase, the DHFR oxidoreductase and
ribonuclease A (RNaseA).

For each member, extensive molecular dynamics simulations were carried
out. The authors next compared the dynamics-based features that directly
impacted the known rate-limiting step of the enzyme catalytic
activity. This important technical step allowed Ramanathan {\em et al.}
to address in a direct and precise way the functionally-oriented
dynamical aspects of the proteins without relying on their
dynamics-based alignment or considering general aspects of the internal
dynamics that are inconsequential for biological
functionality\cite{Ramanathan:2011:PLoS-Biol:22087074}.

By these means Ramanathan {\em et al.} ascertained that the
reaction-coupled motions of the members of each of the three types of
enzymes were highly similar. Because the members were picked from
different species it was further concluded that the detailed
functionally-oriented dynamical aspects have been evolutionarily
conserved.

The analysis established two further notable features. First, the
dynamical similarities found for the homologous CypA entries were found
to extend to the non-homologous PIN1 peptidyl-prolyl isomerase. In
consideration of the structural differences of the modelled structure of
Pin1 and CypA it was concluded that the reaction-coupled motions of the
enzymes were conserved despite the structural differences. Secondly, it
was observed that the dynamical aspects influencing the functional
activity involved regions that are not necessarily near the active site,
thus pointing out at an overall interplay of local and global aspects in
the functional ``mechanics'' of the enzymes. The fact that these
features might hold for several other enzymes is reinforced by the
consistency with the findings reported earlier for members of the
proteases family as well as by instances such as R67 dihydrofolate
reductase where enzyme flexibility has been argued to impact the
catalysed reaction \cite{Kamath:2010:Biochemistry:20795731}.

\subsection{Comparison of general dynamical patterns in members of the SCOP database}

Besides the above-mentioned studies, a comparative investigation of
mean-square fluctuation profiles and mode shapes was recently undertaken
by Tobi~\cite{Tobi:2012:Proteins:22275069} for an extensive set of
entries from the SCOP/Astral
database\cite{Andreeva:2008:Nucleic-Acids-Res:18000004,Chandonia:2004:Nucleic-Acids-Res:14681391}.
A distinctive point of the analysis of
ref.~\cite{Tobi:2012:Proteins:22275069} is the fact that the set of
amino acids over which the dynamical properties are automatically compared is not
identified by sequence or structural alignments, but by matching the
fluctuation (or mode) amplitude profile itself, as first envisaged by Keskin {\em et al.}\cite{keskin}

A key ingredient of this comparative approach is the use of the isotropic
Gaussian network model\cite{bahar97}.  Because this phenomenological model does not
possess the full rotational-translational invariance of the
three-dimensional elastic networks, its essential dynamical spaces have
a one-dimensional character.  By restricting considerations to the
one-dimensional profile of a single mode (or of the mean-square
fluctuation) Tobi used a dynamics-based programming strategy to identify
corresponding amino acids for various pairs of proteins.

Significant matches were reported for pairs of proteins with different
overall structural organization.  Consistently with the isotropic
character of the elastic network model, the lowest energy mode of these
matching proteins typically exhibited a single node located aproximately
in the middle of the matching subchain, thus entailing a hinge-bending
motion. This motion was prototypically illustrated in
ref.~\cite{Tobi:2012:Proteins:22275069} for two pairs of entries:
OPRTase (PDBid, 1s7o chain A) with Mediator complex subunit 21 (PDBid
1ykh chain A fragment 111-205) and Baseplate wedge protein 9 (PDBid,
1s2e chain A) with transcarboxylase (PDBid 1rqh chain A fragment
307-474).

\begin{figure*}[htp]
\includegraphics[width=0.95\textwidth]{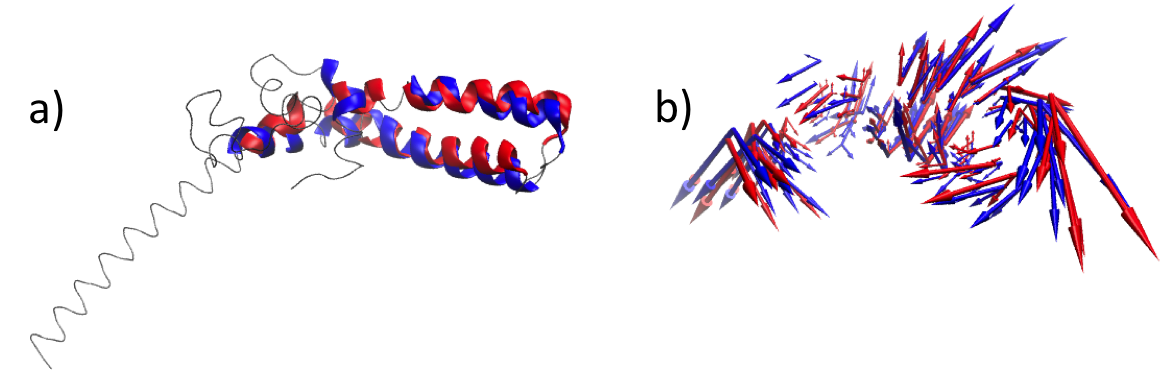}
\caption{Dynamics-based alignment of two OPRTase, (PDBid: 1s7o chain A) and Mediator complex subunit 21 (PDBid: 1ykh chain A)  discussed in ref.~\cite{Tobi:2012:Proteins:22275069}.
The structural superposition of the aligned regions (ribbons) and three best-matching modes (arrows) are shown in panel a and b, respectively. Aligned elements of OPRTase are shown in blue, while those of Mediator comples subunit 21 are shown in red.}
\label{fig:tobi}
\end{figure*}

Notably, the former of these two pairs has also a significant
dynamics-based alignment according to the scheme of Zen. {\em et al.}
which employs a three-dimensional elastic network model as well as the
integration of the dynamics of non-corresponding amino acids). The
corresponding Aladyn alignment is shown in Fig.~\ref{fig:tobi}.

\subsection{Comparison of the structural variability in a protein superfamily with the internal dynamics of its members}

An interesting problem regards the extent to which evolutionary
conformational drifts observed in proteins superfamilies occurs along
the essential dynamical spaces of the family members.

This question was first posed by Leo-Macias {\em et al.}
\cite{LeoMacias:2005:Biophys-J:15542556} who considered 35
representative protein families. For each family, the members were first
structurally aligned to identify the common core and then a principal
component analysis was carried our to obtain the main deformation
modes. The latter were finally compared with the essential dynamical
spaces obtained from elastic network models. {The comparison of the two sets of spaces, which nowadays can be largely automated with the aid of bioinformatic tools such as ProDy\cite{prody}, indicated a good  mutual consistency.}

The investigation of Leo-Macias {\em et al.} was recently extended by
Velazquez-Muriel {\em et
  al.}~\cite{VelazquezMuriel:2009:BMC-Struct-Biol:19220918} who
considered a larger set of 55 families and used atomistic MD
simulations. This study reported that the conformational space explored
in MD simulations at constant-temperature has a smaller breadth than
that spanned by known members of the same superfamily. However, the
complexity of the explored space is significantly larger for MD
simulations than for the internal variability of protein
superfamilies. In this study the complexity was defined and measured as
the minimal number of essential modes required to account for the same
fraction of the global mean-square fluctuation of the superfamily or MD
trajectory.

Based on these findings, Velazquez-Muriel {\em et
  al.}~\cite{VelazquezMuriel:2009:BMC-Struct-Biol:19220918} concluded
that the structural evolution of superfamilies has occurred in diverse
and much richer ways than those kinetically accessible in thermal
equilibrium to any of the superfamily members. Yet, such enhanced
conformational variability was constrained in fewer generalised
directions, compared to those that are {\em a priori} kinetically
accessible.

These conclusions, in turn, prompted the speculation that the
restrictions to the viable superfamily ``conformational complexity''
reflect the evolutionary pressure to preserve certain patterns of
structural fluctuations/motion that cannot be arbitrarily modified
without compromising dynamics-based aspects relevant to function. The
effect was most evident for enzymes, where the largest restrictions of
the conformational variability was
observed~\cite{VelazquezMuriel:2009:BMC-Struct-Biol:19220918}.

The possibility that physics-based constraints may also promote the
consistency of the evolutionary deformation modes and essential dynamical spaces was explored by Echave and coworkers in refs.~\cite{Echave:2010:Proteins:19731380,Echave2012}

\subsection{Dynamics-based alignment of proteins with different structure and function}

We now report on the studies of Zen {\em et al.}
\cite{Zen:2008:Protein-Sci:18369194} who carried out comparisons of the
internal dynamics of a comprehensive set of 76 enzymes covering the six
main functional groups (oxydoreductases, transferases, hydrolases,
lyases, ligases).

The analysis of Zen {\em et al.} was aimed at ascertaining whether similar
functionally-oriented dynamical properties (arising from either
evolutionary conservation or convergence) could be found in enzymes with
major sequence and structure differences. 

\begin{figure}[htp]
\includegraphics[width=0.45\textwidth]{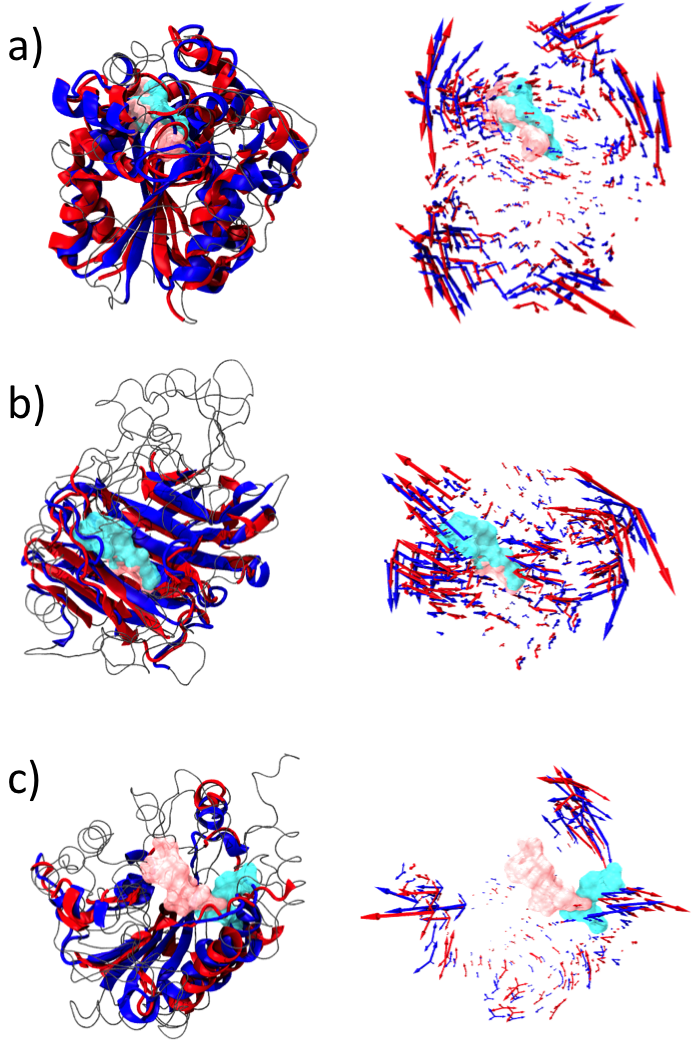}
\caption{Examples of significant dynamics-based alignments of proteins
  with different degree of structural and functional similarities (captured by the CATH code and primary EC number, respectively).
 The examples  are taken from ref.~\cite{Zen:2008:Protein-Sci:18369194} and the
  alignments were produced with the Aladyn web-server. The aligned proteins in panel (a)
 have the same fold (they share the full cath code) but have different function. The pair in panel (b) have the same function but different CATH architecture. The pair in panel (c) differ by CATH architecture and function.
The pair in panel (a) involves a haloalkane dehalogenase (PDBid 2had, CATH: 3.40.50.1820, EC: 4) and a (s)-acetone-cyanohydrin lyase
(PDBid: 1yb7, CATH: 3.40.50.1820, EC: 3). The pair in panel 9b) involves a Cellobiohydrolase i (PDBid: 1dy4, CATH: 2.70.100.10, EC: 3) and a glucanase (PDBid: 2ayh, CATH: 2.60.120.200, EC: 3). The pair in panel (c) involves an exonuclease (PDBid: 1ako, CATH: 3.60.10.10, EC: 3) and an Enoyl-reductase (PDBid: 1d7o, CATH: 3.40.50.720, EC: 1). For each pair we report separately the structural superposition of the aligned regions (ribbons) and of the top three best-matching modes (arrows). Aligned elements are shown in blue for the first entry of the pair and in red for the second. The active sites are shown in cyan and pink for the first and second entry of the pair, respectively.}
\label{fig:lesk}
\end{figure}

The study entailed the dynamics-based alignment (in the spirit of
section \ref{sec:aladyn}) of all the possible pairings of such
enzymes. About 30 of such pairings were singled out as being outstanding
for statistical significance. Two thirds of such pairings involved
enzymes with detectable sequence homology or structural similarity as
resulting by global or partial structural superposition using the DALI
alignment program.  One such example is offered by the pair 1yb7-2had
which share the full CATH code, despite the different function. The
dynamics-based alignment of this pair is shown in
Fig.~\ref{fig:lesk}a where one can observe the remarkable structural superposition of the
molecules' active sites.

Interestingly, the remaining third of the significant pairings involved
entries whose structural relatedness was not significant by standard
alignment criteria and occasionally involved enzymes with different
function, i.e. different primary Enzyme Commission (EC) number.

Two such pairs are respectively, 1dy4-2ayh and 1ako-1d7o, which are
respectively shown in panels b and c of Fig.~\ref{fig:lesk}.  It is seen
that while the overall structural correspondence is limited (and in fact
aligned regions can have different secondary structure content), the alignment
reflects a very good consistency of the matching modes as well as the
superposition of the known active regions.

As for the previously discussed case of proteases, the match of the
latter and the fact that the matching modes entail the modulation of the
region surrounding the active site, support the notion that common 
functionally-oriented  dynamics-based properties can be detected in proteins
that possibly differ by structure and even detailed catalytic chemistry\cite{Ramanathan:2011:PLoS-Biol:22087074,Hensen:2012:PLoS-One:22606222}.

\subsection{Comparing large-scale movements of multidomain proteins}
\label{sec:multidom}

As anticipated at the end of section \ref{sec:ed}, a particularly challenging
case for characterizing protein internal dynamics, as well as comparing
it, is represented by proteins comprising mobile domains.

For such molecules, in fact, the relative displacements of the mobile
domains can be so large that the motion is only poorly described by
linearly superimposing a few essential modes onto a reference structure,
see Fig.~3 in ref.~\cite{Song:2006:Proteins:16447281}. A familiar
example is offered by the opening of a door: the larger the opening
angle, the poorer the directional consistency of the initial
displacement of the door's edge and the difference vector of the initial
and final edge positions. As a consequence, the essential dynamical
spaces calculated for a short trajectory, or by applying elastic network
models on a specific protein conformer, can only limitedly capture and
describe large-amplitude motions in such complexes.  Furthermore, the
very same calculation of essential dynamical spaces from extensive MD
simulations can be problematic because they rely on the use of
rigid-structural alignments which cannot well superimpose the visited
conformers over all their amino acids.

At least for some proteins with mobile subdomains, using internal
angular coordinates instead of Cartesian displacements can provide a
viable alternative for describing the large-amplitude protein
motion\cite{Mendez:2010:Phys-Rev-Lett:20867208,Orozco:2011:Adv-Protein-Chem-Struct-Biol:21920324,LopezBlanco:2011:Bioinformatics:21873636}.

\begin{figure*}[htp]
\includegraphics[width=0.95\textwidth]{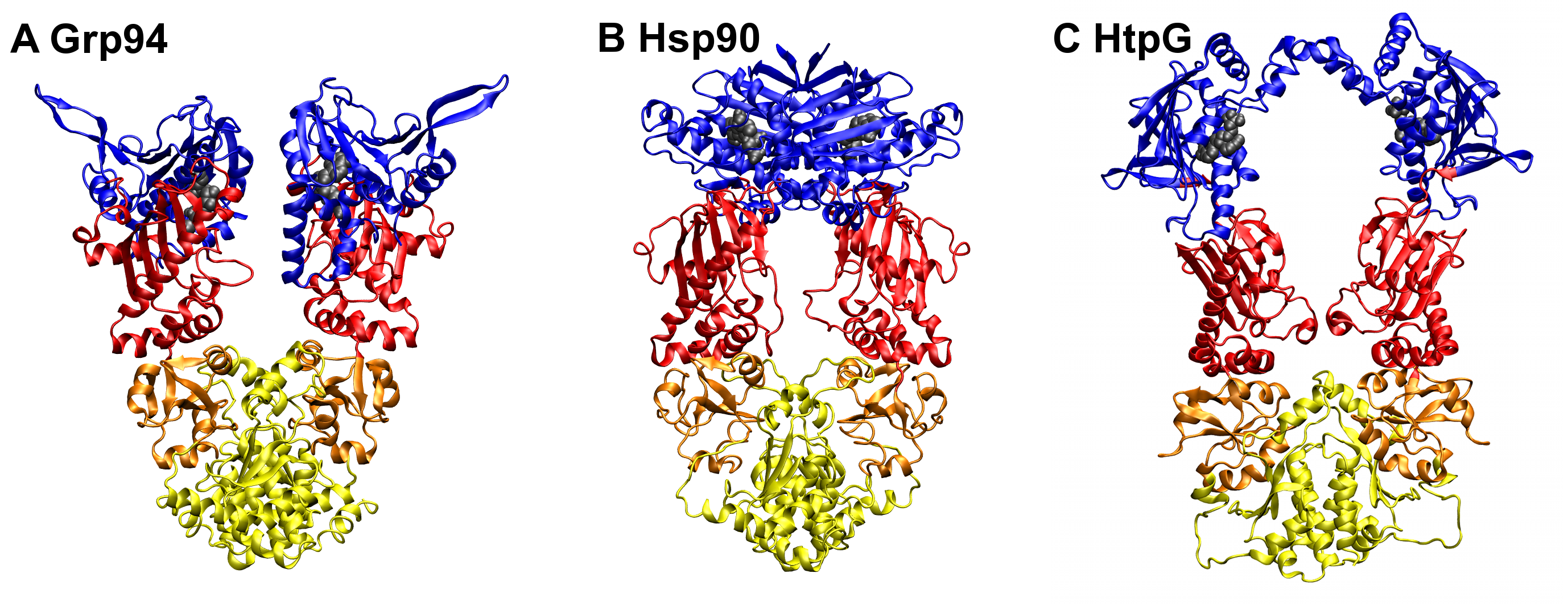}
\caption{Crystallographic structures of three HSP90 conformers used in
  the comparative dynamics study of
  ref.~\cite{Morra:2012:PLoS-Comput-Biol:22457611}. The structures
  correspond to: (A) canine ATP-bound Grp94 structure, PDBid: 2o1u; (B)
  yeast ATP-bound Hsp90 structure, PDBid: 2cg9; (C) HtpG structure,
  PDBid 2iop.pdb. Different colors are used to highlight the various structural subdomains:
  blue, N-terminal domains; Red, M-large domains; Orange, M-small
  domains; Yellow, C-terminal domains. Reproduced from Fig.~1 of
  ref.~\cite{Morra:2012:PLoS-Comput-Biol:22457611}.}
\label{fig:hsp90}
\end{figure*}

The fact that suitably-defined angular coordinates can be used for
comparing the dynamics of proteins articulated in several domains was
recently illustrated by Morra {\em et al.} in
ref.~\cite{Morra:2012:PLoS-Comput-Biol:22457611}.  This study considered
three homodimeric multidomain HSP90 chaperones, namely mammalian Grp94,
yeast Hsp90 and {\em E.coli} HtpG. The three chaperones, which are
represented in Fig.~\ref{fig:hsp90}, have a mutual sequence identity of
$\sim$45\% and most of their amino acids can be put into one-to-one
correspondence by using flexible structural
alignment\cite{Ye:2004:Nucleic-Acids-Res:15215455}.

The internal dynamics of the chaperones was characterised by
extensive molecular dynamics simulations started from different initial
conformers which differed by the presence and type of bound
ligand. Next, to extract the large-scale dynamical features that are
shared by the chaperones, considerations were restricted to the
extensive set of corresponding amino acids. 

The motion of such set was found to be well approximated by the relative
rigid-like movements of three quasi-rigid domains (similar, but not
equal, to the structural ones). As a matter of fact, for all three
chaperones it was possible to identify two consensus hinges and axes of
motion controlling the rotation of the side-domains relative to the core
of each protomer. Notably, one of the hinges (the one at the boundary of
the N-terminal and Middle domain) occurs in correspondence of a site
that had been previously shown to be important to chaperone
functionality. In fact, it was validated as a as a potential target for
HSP90 inhibition\cite{Tsutsumi:2009:Nat-Struct-Mol-Biol:19838189,Vasko:2010:ACS-Med-Chem-Lett:20730035}.  Based on the detailed analysis of the same simulations carried out in
ref.~\cite{Morra:2012:PLoS-Comput-Biol:22457611} it was further
concluded that an analogous role could be played by the site
accommodating the second hinge.

The study of ref.~\cite{Morra:2012:PLoS-Comput-Biol:22457611} therefore
suggests that comparative dynamical analysis based on quasi-rigid
protein domain movements could represent a promising avenue for
identifying functional relationships in multidomain proteins and possibly protein complexes too.

\subsection{A dynamics-based metric for protein space}

We conclude the overview by reporting on the recent work of Hensen {\em
  et al.}~\cite{Hensen:2012:PLoS-One:22606222} who considered a set of
$\sim 100$ proteins covering the main known folds and compared their
structural features and especially a comprehensive series of dynamical
observables calculated from 100-ns long atomistic MD simulations.  In
particular, to each protein entry, Hensen {\em et al.}  associated a
dynamical ``fingerprint'' consisting of a multidimensional array whose
components were dynamics-based scalars. These scalar quantities included
the spread of the essential dynamics eivenvalues, the roughness of the
free energy landscape, the root-mean-square-deviation from the
crystallographic structure, the root-mean-square fluctuations from the
average structure etc.

At variance with the studies mentioned earlier, which aimed at detecting
detailed dynamical correspondences among proteins, the investigation of
ref.~\cite{Hensen:2012:PLoS-One:22606222} was mostly targeted to
establishing the overall features of the space spanned by the dynamical
fingerprints. In particular, Hensen {\em et al.} meant to introduce a
dynamics-based metric to explore the occupation of the fingerprint space
(termed the ``dynasome space'') and understand e.g. whether structurally
or functionally similar proteins  can be clustered.

From this survey, the authors concluded that in the considered dynamical
space, proteins are not partitioned in distinct clusters but are
distributed rather continuously. This interesting aspect therefore
parallels the findings of recent studies which support the view that
structural properties cover a continuum rather than a discrete
succession of conformers
\cite{Zhang:2006:Proc-Natl-Acad-Sci-U-S-A:16478803,Skolnick:2009:Proc-Natl-Acad-Sci-U-S-A:19805219,Xie:2008:Proc-Natl-Acad-Sci-U-S-A:18385384,PascualGarcia:2009:PLoS-Comput-Biol:19325884}.

The analysis has further revealed the strong connection between
dynamical and structural similarities, consistently with the studies,
mentioned earlier in this review, where the structural relatedness has
been frequently associated to strong dynamical implications.

It is interesting to observe that, as in the study of Zen {\em et al.}
\cite{Zen:2008:Protein-Sci:18369194} described in the previous section,
the analysis of Hensen {\em et al.} \cite{Hensen:2012:PLoS-One:22606222}
has highlighted the possible existence of appreciable dynamical
similarities in proteins with limited structural relatedness. The
example offered by the authors pertained to the pairing of two
hydrolases, serralysin and rhizopuspepsin (PDB codes 1sat and
2apr). Their structural alignment is non-significant according to DALI
statistical criteria while in the dynamic metric space considered by the
authors they have a strong dynamical proximity. Consistently with this
finding the Aladyn alignment of this pair, which involves 79 amino
acids) is statistically significant too as the observed RMSIP=0.66 and
the associated p-value is 0.025.

Finally, by examining the dynamic fingerprint of functionally-related
proteins Hensen {\em et al.}\cite{Hensen:2012:PLoS-One:22606222} concluded that it ought to be possible to
reliably establish and assign proteins function based on their
neighbours in the metric dynamic space.  Indeed, the possibility to
carry out functional assignments on the basis of dynamics-based data
represents a very interesting avenue with several practical
ramifications.

As a related issue we report that pairwise
dynamics-based alignments have been previously carried out with the
purpose of predicting the active site of proteins for which standard
homology-based approaches are not applicable.  In particular, this
approach was undertaken to predict the nucleic-acids binding sites of
proteins adopting non-canonical OB-folds, as discussed in
ref.~\cite{Zen:2009:Bioinformatics:19487258}.

\section{Conclusions}

Over the past decades, several bioinformatics tools and computational
methods have been introduced and systematically applied to clarify
aspects of the relationship between structure and dynamics for protein
and enzymes.

Many such studies contributed to clarifying how the interplay of
structure and internal dynamics of various proteins impacts their
biological functionality. The latter, in fact, is often -- though not
always -- associated with the innate capability of these biomolecules to
sustain concerted, large-scale conformational changes so to bind
ligands, change oligomeric state etc.

In recent years, besides the well-established approach of dissecting
such properties for specific, individual proteins and enzymes, there has
been a growing interest for comparative studies of proteins' internal
dynamics.

In such studies, covered by this review, the key dynamics-based properties of
proteins are singled out by identifying those features (such as
essential dynamical spaces, mean square fluctuation profiles, relaxation times etc.)
that are shared by proteins with different degrees of sequence,
structure and functional similarities.

Such comparative investigations have been carried out with two main
purposes: characterizing functionally-oriented mechanisms for specific
groups of proteins and understanding the more general organization of
the ``protein universe'' by complementing the sequence- and structural
prespectives with a dynamics-based one.

For the first objective, detailed comparative tools have been developed,
including the so-called dynamics-based alignments which use
dynamics-based properties to establish one-to-one correspondences of
amino-acids in different proteins. These strategies have been used to
identify common hinge-bending motions in multi-domain proteins, to
complement sequence- and structural-alignment in singling out
functionally-relevant regions in proteins with different degrees of
homologies, and to highlight common large-scale movements in proteins
that differ significantly by fold and/or function.

The latter aspect, is tighly connected to the second objective, namely
the development and use of dynamics-based criteria to trace elusive
evolutionary relationships and group/classify proteins by their internal
dynamics\cite{Gerstein:1994:Biochemistry:8204609,Gerstein:1998:Nucleic-Acids-Res:9722650,Capozzi:2007:J-Proteome-Res:17935310}. This
perspective has been pursued so far to highlight the degree of
conservation of the amplitude of amino acid fluctuations in protein
families and superfamilies, to clarify the extent to which the
structural variations accumulated within protein superfamilies have
occurred along the ``innate'' directions of structural fluctuations of
its members, and even to introduce a metric to quantify how evenly are
proteins distributed in a generalized dynamics-space.  The latter
prespective can have important implications for functional assignment.

In conclusion, the valuable findings provided by the recent introduction
of methods for comparing detailed or general dynamical properties of
proteins suggest that they could be profitably used in conjuction with
classic comparative methods to characterize proteins at the various
steps of the sequence $\to$ structure $\to$ function ladder.

Arguably, the progress towards this goal would be greatly aided by the
development of unsupervised methods to single out those dynamical
features that are more likely attributed to the biological functionality
of a given protein and by the more systematic investigation of
evolutionary relationships from a detailed dynamics-based perspective.

{\bf Acknowledgements.} I am grateful to P.~Agarwal, G.~Bussi, V.~Carnevale, J.~Echave, A. Finkelstein, H. Gruebmueller, R.~Jernigan, A.~Lesk, D.~Tobi and A.~Zen for very valuable discussions. We acknowledge support from the Italian Ministry of Education.

{\em NOTICE: this is the author's version of a work that was accepted for publication in Physics of Life Reviews. Changes resulting from the publishing process, such as peer review, editing, corrections, structural formatting, and other quality control mechanisms may not be reflected in this document. Changes may have been made to this work since it was submitted for publication. A definitive version was subsequently published in Phys of Life Reviews, DOI\# 10.1016/j.plrev.2012.10.009"}

\bibliographystyle{elsarticle-harv}
\bibliography{references}

\end{document}